\documentclass{aa}  
\usepackage{graphicx}
\usepackage{txfonts}
\usepackage{color}
\usepackage{soul}
\begin{document}

\title{
{\it Herschel}/HIFI\thanks{{\it Herschel} is an ESA space observatory with
science instruments provided by European-led Principal Investigator
consortia and with important participation from NASA.  HIFI is the 
{\it Herschel} Heterodyne Instrument for the Far Infrared.}\   
observations of the
circumstellar ammonia lines in IRC$+$10216
}

\author{
M.~R.~Schmidt\inst{1}
\and 
J.~H.~He\inst{2}
\and
R.~Szczerba\inst{1}
\and
V.~Bujarrabal\inst{3}
\and
J.~Alcolea\inst{4}
\and
J.~Cernicharo\inst{5}
\and
L.~Decin\inst{6,7}
\and
K.~Justtanont\inst{8}
\and
D.~Teyssier\inst{9}
\and
K.~M.~Menten\inst{10}
\and
D.~A.~Neufeld\inst{11}
\and
H.~Olofsson\inst{8,12}
\and
P.~Planesas\inst{4}
\and
A.~P.~Marston\inst{9}
\and
A.~M.~Sobolev\inst{13}
\and
A.~de~Koter\inst{7}
\and
F.~L.~Sch\"oier\thanks{Deceased 14 January 2011}\inst{8}
}

\offprints{M. R. Schmidt}

\institute{   
N. Copernicus Astronomical Center, Rabia{\'n}ska 8, 87-100 Toru{\'n}, Poland
\email{schmidt@ncac.torun.pl}
\and
Key Laboratory for the Structure and Evolution of Celestial Objects, Yunnan
Observatories, Chinese Academy of Sciences, P.O.  Box 110, Kunming, Yunnan
Province, China
\and
Observatorio Astron\'omico Nacional. Ap 112, E-28803 
Alcal\'a de Henares, Spain
\and
Observatorio Astron\'omico Nacional (IGN), Alfonso XII N$^{\circ}$3,
              E-28014 Madrid, Spain 
\and
ICMM, CSIC, group of Molecular Astrophysics, C/Sor Juana In\'es de la Cruz N3, 
28049 Cantoblanco (Madrid), Spain
\and
Instituut voor Sterrenkunde,
             Katholieke Universiteit Leuven, Celestijnenlaan 200D, 3001
Leuven, Belgium
\and
    Sterrenkundig Instituut Anton Pannekoek, University of Amsterdam,
Science Park 904, NL-1098 Amsterdam, The Netherlands
\and
Chalmers University of Technology, Department of Earth and Space Sciences, 
Onsala Space Observatory, S-439 92 Onsala, Sweden
\and
European Space Astronomy Centre, ESA, P.O. Box 78, E-28691
Villanueva de la Ca\~nada, Madrid, Spain
\and
Max-Planck-Institut f{\"u}r Radioastronomie, Auf dem H{\"u}gel 69,
D-53121 Bonn, Germany 
\and
Johns Hopkins Universtity, Baltimore, MD 21218, USA
\and
Department of Astronomy, AlbaNova University Center, Stockholm  
University, SE--10691 Stockholm, Sweden
\and
Ural Federal University, Astronomical Observatory, 620000 Ekaterinburg, Russian Federation
}

\date{Received; accepted}
\abstract
{
A discrepancy exists between the abundance of ammonia (NH$_3$) 
derived previously for the
circumstellar envelope (CSE) of IRC$+$10216 from far-IR 
submillimeter rotational lines and that inferred from radio inversion or
mid-infrared (MIR) absorption transitions.
}
{
To address the discrepancy described above,
new high-resolution far-infrared (FIR) observations of both ortho- and
para-NH$_3$ transitions toward IRC$+$10216 were obtained with {\it
Herschel}, with the goal of determining the ammonia abundance and constraining
the distribution of NH$_3$ in the envelope of  IRC$+$10216.
}
 { 
We used the Heterodyne Instrument for the Far Infrared (HIFI) on board 
{\it Herschel} to observe all rotational transitions up to the $J=3$ level 
(three ortho- and six para-NH$_3$ lines).
We conducted non-LTE multilevel radiative transfer modelling, including the effects of
near-infrared (NIR) radiative pumping through vibrational transitions.  The
computed emission line profiles are compared with the new HIFI data, the
radio inversion transitions, and the MIR absorption lines in the $\nu_2$
band taken from the literature.
}
 {
We found  that NIR pumping is of key importance for understanding the
excitation of rotational levels of NH$_3$.  The derived NH$_3$ abundances
relative to molecular hydrogen were $(2.8\pm 0.5)\times10^{-8}$ for
ortho-NH$_3$ and $(3.2^{+0.7}_{-0.6})\times10^{-8}$ for para-NH$_3$, consistent
with an ortho/para ratio of 1.  These values are in a rough agreement with
abundances derived from the inversion transitions, as well as with the total
abundance of NH$_3$ inferred from the MIR absorption lines.  To explain the
observed rotational transitions, ammonia must be formed near to the central
star at a radius close to the end of the wind acceleration region, but no larger
than about 20 stellar radii (1$\sigma$ confidence level).
}
{}
\titlerunning{NH$_3$ in C-rich AGB star IRC$+$10216}
\authorrunning{Schmidt et al.}
\maketitle

\section{Introduction}

Ammonia (NH$_3$) was the first polyatomic molecule discovered in space
\citep{Cheung1968}, and has subsequently been one of the most extensively
observed molecules in the interstellar medium \citep{Ho1983}, with a set 23
GHz of inversion transitions that are readily detectable from many radio
telescopes.  Emission lines resulting from these transitions are widely
detected in the dense parts of both dark and star-forming molecular clouds
\citep[see e.g.][]{Harju1993,Jijina1999,Wienen2012}.  Because ammonia is a
symmetric top molecule, an analysis of its excitation allows the effects of
temperature and density to be determined separately; as a result, the
metastable lines of ammonia are often used to derive the temperature and
density of dense clumps in molecular clouds. 
\citep{Walmsley1983,Danby1988,Kirsanova2014}.

Ammonia has also been detected in the circumstellar envelopes of evolved
stars, but has been observed less extensively in such environments than
in the interstellar medium.  Several absorption features in the $\nu_2$
vibration$-$rotational bands around 10\,$\mu$m were detected in some
asymptotic giant branch (AGB) stars, as well as in a few massive supergiants
\citep[][]{Betz1979,McLaren1980,Betz:1987lr}.  At about the same time, the
1.3 cm wavelength inversion transitions of ammonia were detected from C-rich
AGB and post-AGB stars \citep[][]{Betz:1987lr,
Nguyen-Q-Rieu1984,Nguyen-Q-Rieu:1986th}.

In IRC$+$10216 (CW\,Leo), ammonia was observed for the first time by its
infrared absorption lines in the $\nu_2$ band around 10\,$\mu$m
\citep{Betz1979}, and its detection in radio inversion transitions was
announced shortly thereafter \citep{Bell:1980hc}.  In the following years,
new observations of IRC$+$10216 were performed in both infrared absorption
\citep{Keady1993,Monnier2000b} and radio inversion lines
\citep{Kwok1981,Bell1982,Nguyen-Q-Rieu1984,Gong2015}.

IRC$+$10216 was detected in the Two-micron Sky Survey as the brightest
infrared sky object at 5\,$\mu$m outside the solar system
\citep{Becklin1969}, and soon its heavily obscured central star was
identified as being C-rich based on the presence of strong CN absorption
bands \citep{Herbig1970,Miller1970}.  Presently, IRC$+$10216 is the best
studied C-rich AGB star.  About half of the molecules detected in space are
seen in the envelope of this source, due to its proximity (distance, $d$, of
130~pc, see \citep{Menten2012}) and a relatively large mass loss rate
$\sim 1-4 \times10^{-5}$ M$_{\sun}$ yr$^{-1}$
\citep[e.g.][]{Groenewegen:1998rw,Truong-Bach:1991qa}.

The development of heterodyne technology for observations in the
far-infrared (FIR) enabled the first detection of the 1$_0$(s)-0$_0$(a)
rotational transition of ortho-NH$_3$ by the {\it Odin} satellite toward
IRC$+$10216 \citep{Hasegawa2006}, which was followed by the detection of the
same transition in some O-rich AGB stars and red-supergiants
\citep{Menten:2010pd} by the Heterodyne Instrument for the Far Infrared
\citep[{HIFI},][]{de-Graauw:2010qe} on board the 
{\it Herschel} 
Space Observatory
\citep[{\it Herschel},][]{Pilbratt:2010oq}.  Furthermore, by fully
exploiting the capabilities of the HIFI instrument, we were able to observe
all nine rotational transitions up to the $J=3$ levels of ortho- and
para-NH$_3$ (this paper) in the envelope of IRC$+$10216.

There is a clear discrepancy between the estimated amount of para-NH$_3$
relative to molecular hydrogen in IRC$+$10216 derived from radio inversion
lines \citep[$3\times10^{-8}$;][]{Kwok1981} and that of ortho-NH$_3$ derived
from its lowest rotational line \citep[about $10^{-6}$;][]{Hasegawa2006}. 
Nevertheless, both values, which lie within the range of ammonia abundances
observed towards other C-rich and O-rich stars, and cannot be explained by
standard chemical models \citep[see discussion in][]{Menten:2010pd}. 
However, the chemical model presented recently by \cite{Decin:2010kl} seems
able to reproduce the high abundance of ammonia observed in IRC$+$10216 (see
their Fig.\,3).  Their model, constructed with the aim of explaining the
presence of water vapour in the envelope of this C-rich star, assumes a
clumpy envelope structure, where a fraction of the interstellar UV photons
is able to penetrate deep into the envelope and dissociate mostly $^{13}$CO,
providing oxygen for O-rich chemistry in the inner warm parts of a C-rich
circumstellar envelope.  On the other hand, \cite{Neufeld:2013fk} have
suggested, on the basis of H$_2$O isotopic ratios, that a recent model
invoking shock chemistry \citep{Cherchneff2012} may provide a more
successful explanation for the presence of water in envelopes of C-rich
stars.  However, this model does not provide information on the ammonia
formation.

In this paper we present new {\it Herschel}/HIFI observations of nine
rotational transitions of NH$_3$ (three ortho and six para lines), eight of
which have been detected for the first time, and an analysis of their
implications with the use of detailed modelling.

The observations and data reduction are presented in Section 2.  Section 3
is devoted to the description of the molecular structure of ammonia, while
Section 4 presents details of the modelling procedure and the best fits found
for all rotational transitions.  In Section 5 we discuss the results thereby
obtained and their consequences.  Finally, a summary of this study is
presented in Section 6.

\section{Observations}

\begin{figure}
\centering
\includegraphics[width=8.5cm]{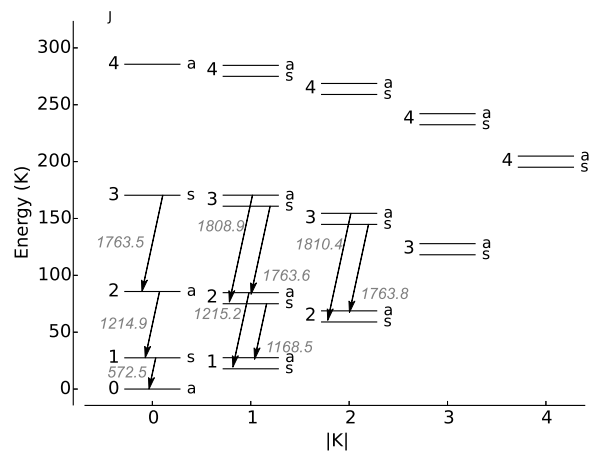}
\caption{
Diagram of energy levels of ortho- (|$K$|=0,3, etc) and para-NH$_{3}$ (|$K$|=1,2,4, etc).
The inversion splitting between levels of different symmetry is 
exaggerated for clarity. The observed rotational transitions
with frequencies in GHz are indicated with arrows.
}
\label{fignh3}
\end{figure}

\begin{table*}
\caption{Summary of NH$_3$ observations in IRC$+$10216.}
\label{TableObs}
\begin{tabular}{lrcrrccccrl}
\hline\hline
Transition\tablefootmark{a}      &  Frequency & Band &
E$_u$\tablefootmark{b} & {\it Herschel} & Obs. date   &
Phase\tablefootmark{c} & 
HPBW      & 
$\eta_{mb}$  & 
Int. flux     &  Observing  \\
                   &   (GHz)    &     & (K) &  OBSID       &             &       &    &      &(K\,km\,s$^{-1}$) & mode \\
\hline
{\bf 1$_0$(s)-0$_0$(a)}  &   {\bf 572.498}  &  1b &  29 &  1342195794  &  2010-05-04 & 0.22  & 37.5\arcsec &  0.62  & 22.7$\pm$1.2 & single point \\
                   &            &     &     &  1342196413  &  2010-05-11 & 0.22  &      &             & & single point  \\
2$_1$(s)-1$_1$(a)  &  1168.452  &  5a &  58 &  1342196514  &  2010-05-13 & 0.23  & 18.2\arcsec &  0.59  & 19.5$\pm$2.0 & spectral scan \\
{\bf 2$_0$(a)-1$_0$(s)}  &  {\bf 1214.853}  &  5a &  86 &  1342196514  &  2010-05-13 & 0.23  & 17.5\arcsec &        & 39.2$\pm$3.9 & spectral scan\\  
2$_1$(a)-1$_1$(s)  &  1215.246  &  5a &  60 &  1342196514  &  2010-05-13 & 0.23  & 17.5\arcsec &        & 21.8$\pm$2.2 & spectral scan \\
{\bf 3$_0$(s)-2$_0$(a)}  &  {\bf 1763.524}  &  7a & 170 &  1342233281  &  2011-11-28 & 0.13  & 12.0\arcsec &  0.58  & 48.5$\pm$7.2 & single point \\
3$_1$(s)-2$_1$(a)  &  1763.601  &  7a & 143 &  1342233281  &  2011-11-28 & 0.13  & 12.0\arcsec &        & 23.7$\pm$3.6 & single point \\  
3$_2$(s)-2$_2$(a)  &  1763.823  &  7a & 127 &  1342233281  &  2011-11-28 & 0.13  & 12.0\arcsec &        & 23.2$\pm$2.3 & single point \\  
3$_1$(a)-2$_1$(s)  &  1808.935  &  7b & 144 &  1342196574  &  2010-05-15 & 0.23  & 11.7\arcsec &  0.58  & 18.0$\pm$9.0 & spectral scan \\
3$_2$(a)-2$_2$(s)  &  1810.380  &  7b & 128 &  1342196574  &  2010-05-15 & 0.23  & 11.7\arcsec &        & 18.1$\pm$9.0 & spectral scan  \\  
\hline
\hline
\end{tabular}
\tablefoot{
\tablefoottext{a}{ortho transitions and their frequencies are indicated in bold
face.}
\tablefoottext{b}{energies of levels for para-NH$_3$ should be increased by
22\,K if put on the common energy scale with ortho-NH$_3$.}
\tablefoottext{c}{
counted from the reference maximum phase of the light curve,  $\phi$=0 on Julian date JD=2454554, 
and period of 630 days \citep{Menten2012}.}
}
\end{table*}

Observations of IRC$+$10216 were carried out with the {\it Herschel}/HIFI
instrument as part of the HIFISTARS Guaranteed Time Key Program (Proposal
Id: KPGT\_vbujarra\_1; PI: V.  Bujarrabal); in addition, a spectral line
survey has been carried out 
for 
this object (Proposal Id:
GT1{\_}jcernich\_4; PI: J.  Cernicharo).  The observed rotational lines of
ortho- and para-NH$_3$ are listed in Table~\ref{TableObs} and indicated in
Fig.\,\ref{fignh3}, which shows the diagram of the lowest rotational levels
of ammonia.  For each observed transition, Table~\ref{TableObs} gives the
line identification, its frequency in GHz, the corresponding HIFI band, the
energy of the upper level in K, the {\it Herschel} OBServation ID, the date
of the observation, the optical phase ($\phi$) of the observation (counted
from the reference maximum phase of the light curve, $\phi$=0, on Julian
date JD=2454554 with an assumed period of 630 days \citep{Menten2012}), the
half power beam width (HPBW) of the {\it Herschel} telescope at the observed
frequency, the main beam efficiency $\eta_{mb}$, the integrated flux in
K\,km\,s$^{-1}$ with its estimated uncertainty, and the observing mode:
single tuning or spectral scan.  In this paper, we analyse wide band
spectrometer (WBS) data only.

The HIFI observations of IRC$+$10216 were reduced with the latest version of
HIPE (13.0) with data processed with Standard Product Generator  (SPG) version 13.0.0.  The data were processed using the HIPE pipeline to Level 2,
which was set to provide intensities expressed as the main beam temperature. 
Main beam efficiencies were taken from the recent measurements
by Mueller et al.
2014\footnote{\path{http://herschel.esac.esa.int/twiki/pub/Public/HifiCalibrationWeb/HifiBeamReleaseNote_Sep2014.pdf}}.
They differ by up to 20\,\% from the older determinations 
by \cite{Roelfsema2012}.  
Frequencies are always given in the frame of the local standard of rest (LSR). 

The statistical uncertainties in the integrated line fluxes, as derived
formally from the r.m.s.\ noise in the spectra (see below) are relatively
small, while it is known that systematic uncertainties in the HIFI flux
calibration are much larger \citep{Roelfsema2012}.  Therefore, somewhat
arbitrarily, we have assumed a 5\% uncertainty in the integrated line flux
for lines observed in Band 1b, a 10\% uncertainty for lines in bands 5a and
7a, and a 50\% uncertainty for lines in band 7b.  However, since two of the
lines observed in band 7a are blended (see below), we increased the assumed
uncertainty in their integrated flux from 10 to 15\%.  The method used to
deconvolve line blends and to derive individual line fluxes is described in
Sect.\,2.2.  The observed line profiles are shown by the solid lines in
Figure~\ref{figb} .

\subsection{Single point observations}

The HIFISTARS observations were all performed in the dual beam switch (DBS)
mode.  In this mode, the HIFI internal steering mirror chops between the
source position and two positions located 3\arcmin\ on either side of the
science target.  There are two entries in Table 1 for the ground-state transition
1$_0$(s)-0$_0$(a) of ortho-NH$_3$, since these two observations were made
with slightly shifted local oscillator frequencies.  This procedure was adopted to
confirm the assignment of any observed spectral feature to either the upper
or lower sideband of the HIFI receivers \citep{Neufeld2010}.  The resultant
spectra were co-added since no contamination was found to originate from the other
side band.  Spectra obtained for the horizontal (H) and vertical (V)
polarizations were found to be very similar (differences smaller than 5\,\%)
and were co-added too.  After resampling to a channel width of 1 km s$^{-1}$, 
the final baseline r.m.s. noise in the coadded spectra was 6 mK in 
band 1b and 40 mK in band 7a.

\subsection{Spectral scan observations}

We made use of data from the full spectral scan of IRC$+$10216
(GT1{\_}jcernich\_4 by Cernicharo et al., 
in preparation) 
to obtain
measurements of five additional transitions of ammonia observed in two
bands, 5a and 7b (see Table~\ref{TableObs}).  The spectral scan observing
mode allows for the unique reconstruction of the single-sideband spectrum in
a wide spectral range at the price of significantly lower signal-to-noise
ratio.  Spectral scan observations were also made  in dual beam switch mode
with fast chop.  Both the spectral resolution and signal-to-noise ratio are
lower for these observations than for the single point ones.  Lines observed
in the scan mode were deconvolved to obtain a single-sideband spectrum.  To
reduce the noise in the profiles, the final spectra were resampled to 1 km
s$^{-1}$ in band 5a and 2 km s$^{-1}$ in band 7b.  The final r.m.s noise in
the resampled spectra was 125 mK and 200 mK in bands 5a and 7b,
respectively.

The para-NH${_3}$ transition 3$_1$(s)-2$_1$(a) at 1763.601 GHz (middle
panel) overlaps the ortho-NH${_3}$ transition 3$_0$(s)-2$_0$(a) at
1763.524 GHz (upper left panel).  Therefore, the shape of the former line was
approximated in its blueshifted part using, as a template, a properly
rescaled profile of the para-NH$_{3}$ 3$_2$(s)-2$_2$(a) emission line (middle
right panel), which was observed with the same beam size and has a
similar excitation.  The line profile of the ortho-NH${_3}$ transition
3$_0$(s)-2$_0$(a) was then obtained by subtracting  the derived
para-NH$_{3}$ 3$_1$(s)-2$_1$(a) line emission.  The resulting profiles are
shown by dot-dashed lines in Fig.\,2.

\subsection{Other observations}

To verify our best fit models, we used radio and mid-infrared data for
ammonia available from the literature.  In this paper, we exploited radio
data from \cite{Gong2015}, who recently observed five metastable inversion
transitions ($J,K$) = (1,1), (2,2), (3,3), (4,4), (6,6) of ortho- and para-
ammonia in a 1.3\,cm line survey toward IRC$+$10216.  The dates of those
observations correspond to phases, $\phi$, from about $\sim$0.62 to
$\sim$0.97 \citep{Gong2015}.  The half power beam width amounts to
40\arcsec\ at the frequency of the inversion lines (about 24 GHz) for
observations with the Effelsberg 100\,m radio telescope \citep{Gong2015}. 
Flux densities, $S_{\nu}$ in Jy, were converted to main beam temperature
$T_{\rm mb}$, in K, using the conversion factors suggested by the authors:
$T_{\rm mb}/S_{\nu}=1.33$ K/Jy for the (6,6) line at 25 GHz and $T_{\rm
mb}/S_{\nu}=1.5$ K/Jy for the remaining lines.
 
By way of comparison with the mid-infrared absorptions of ammonia, we used
profiles of the $\nu_2$ transitions published by \cite{Keady1993} that were
observed at a spectral resolution of 0.009\,cm$^{-1}$ (corresponding to 2.9
km~s$^{-1}$) with the Fourier Transform Spectrometer at the Coud{\'e} focus
of the 3.8\,m Kitt Peak Mayall telescope.  Observations of absorption from
the metastable rotational levels within the ground vibrational level of
ortho- and para- ammonia, $aR(0,0)$, $aQ(2,2)$, $aQ(3,3)$, $aQ(4,4)$,
$aQ(6,6)$ were also used for comparison with the results of our modelling. 
Formally, from the aperture of the telescope, we can estimate the HPBW to be $0\farcs67$, using the expression
$1.22\times\lambda/D$, where $\lambda$ is wavelength and $D$ is diameter of
the telescope aperture.  However, taking into account all instrumental
effects, we estimate that the HPBW, in this case, could be as large as
1\arcsec.

\begin{figure*}
\centering
\includegraphics[width=17cm]{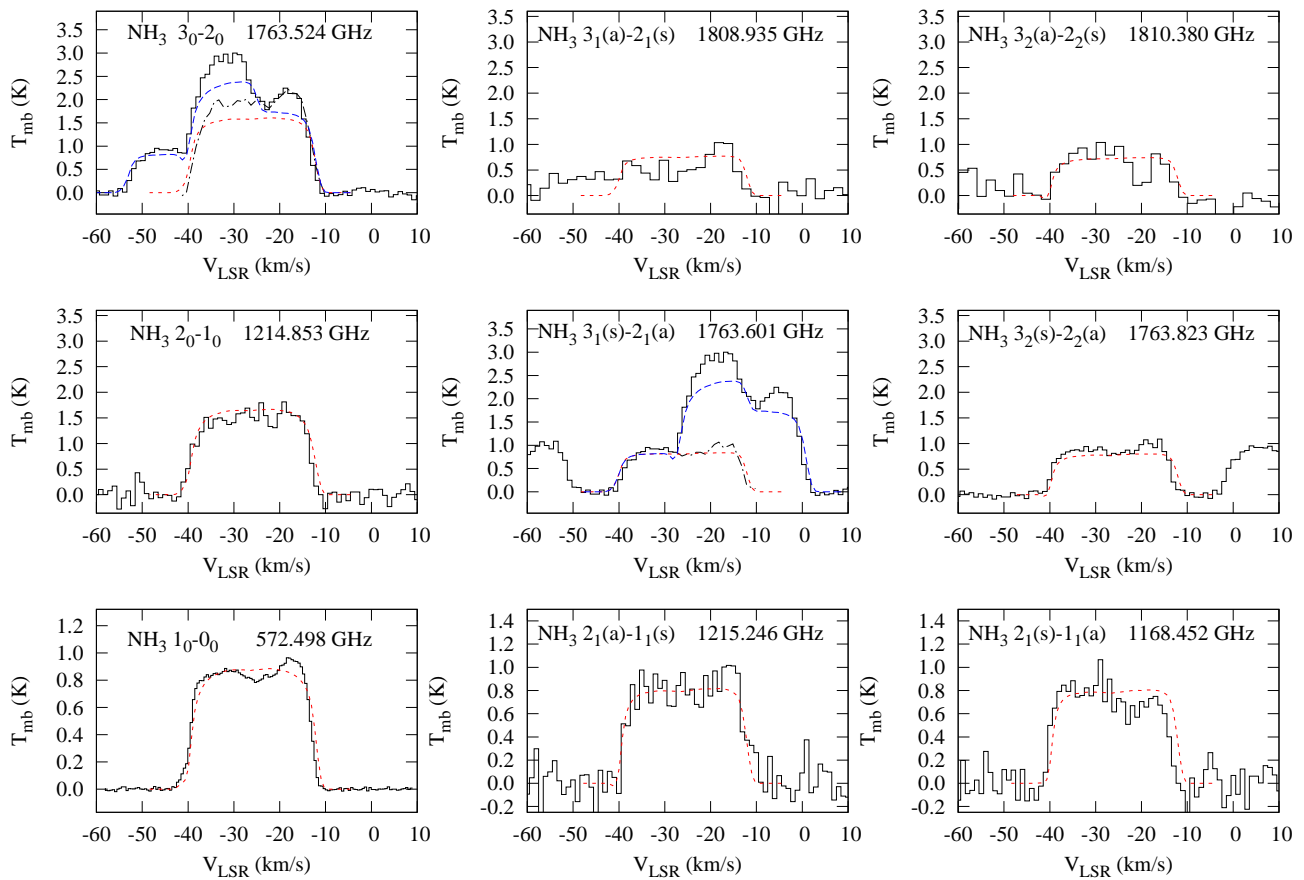}
\caption{
HIFI observations of rotational transitions of ortho-NH$_{3}$ (left column)
and para-NH$_{3}$ (middle and right columns) are shown by solid lines. 
Approximated line profiles (see Sect.\,2.2) of two blended lines
3$_0$(s)-2$_0$(a) and $3_1$(s)-$2_1$(a) are shown by dot-dashed lines. 
Emission profiles are overplotted with theoretical profiles (red dashed
lines) 
from our best fit models computed separately for each ammonia spin
isomer.  The effect of line overlapping in case of the two blends is shown with
blue dashed lines.  The theoretical profiles for the three lines that were
observed at phase $\phi=$\,0.13 (see Table\,1) are rescaled up to mimic
computations at  phase  $\phi=$\,0.23 (see Sect\,4.5 for details).  Our
approach in searching for the best fit is described in Sect.\,4.5, and the best
fitting parameters are compiled in Table\,2.
}
\label{figb}
\end{figure*}

\section{Ammonia model}

Different relative orientations of the spins of the three hydrogen nuclei
give rise to two distinct species of NH$_3$: ortho and para.  The general
criterion governing which levels belong to ortho- and which to para-ammonia
is formulated in terms of representations of the molecular symmetry group
\citep[see][]{BunkerJensen}.  According to this formalism, ortho states
belong to the A$_2$' and A$_2$" representations of the inversional symmetry
group D$_{3h}$, and para states to the E' and E" representations.  For the
electronic ground state of ammonia, this translates into the rule that
ortho-NH$_3$ has levels with $K=3n$, where $n=0,1$,.., and para-NH$_3$ has
all other levels.  The two species do not interact either radiatively or
collisionally.  The ortho- to para-NH$_3$ ratio is determined at the moment
of the formation of the molecule.  Hence, we consider ortho- and
para-ammonia as separate molecular species.  Transitions of the two species
whose frequencies overlap, such as the ortho-NH$_3$ 3$_0$(s)-2$_0$(a)
transition and the para-NH$_3$ 3$_1$(s)-2$_1$(a) transition, open the
possibility of an interaction through the emission and absorption of line
photons.  This is considered further in the discussion below.

Ammonia may oscillate in six vibrational modes: the symmetric stretch
$\nu_1$, symmetric bend $\nu_2$, doubly-degenerate asymmetric stretch
$\nu_3^{l_3}$, and the doubly-degenerate asymmetric bend $\nu_4^{l_4}$. 
Transitions from the ground vibrational state to each of these excited
states are observed as vibrational bands at 3, 10, 2.9, and 6 $\mu$m,
respectively.  The band intensities are characterized by the vibrational
transition moments 0.027, 0.24, 0.018, and 0.083, respectively
\citep{Yurchenko2011}.  As long as we consider only the lowest excited
vibrational modes, the rule governing the assignment of levels of given $K$
to the ortho and para species, discussed above for the ground state, is
preserved for the rotational levels of the symmetric modes and exchanged for
the asymmetric ones.

The vibrational ground state of ammonia is split into two states of opposite
parities, a consequence of the low energy barrier to inversion of the
molecule.  Symmetry considerations exclude half of the rotational levels for
the $K=0$ ladder.  Only transitions with $\Delta K=0$ are electric
dipole-allowed.  As a result, there is a characteristic doubling of
rotational transitions allowed between levels of opposite parities for
transitions with $K\neq 0$.  Forbidden transitions ($\Delta K\neq 0$) are
significantly weaker.  Transitions between split sub-levels are allowed as
well and give rise to a large number of lines around 23-25\,GHz, the so
called inversion lines.  Their hyperfine structure (hfs) splitting may be
neglected in the rotational and vibrational transitions but does influence
the line profiles for inversion lines although for IRC+10216, the maximal
spread of the hfs components is smaller than the CSE's expansion velocity. 
Moreover, the intensities of hfs satellite features relative to that of 
the main
feature (at the central frequency) only become appreciable if an inversion
transition (with possible exception of the (1,1) transition) attains
significant optical depth.

The list of transitions and their strengths was extracted from the recent
BYTe computations \citep{Yurchenko2011}.  In the initial analysis, we
extended the sets of molecular data for ortho- and para-NH$_3$ compiled in
the LAMDA database \citep{Schoier2005} by adding energy levels for the excited
vibrational states.  Two data sets were explored, the first one with only the
$\nu_2$=1 levels,
and the second one with $\nu_1$=1, $\nu_2$=1 and 2, $\nu_3$=1, and $\nu_4$=1 levels. 
The rotational quantum numbers of the vibrational levels were limited to those
originally used in the ground state, i.e.  $J,K\le$7,7 in ortho- and
$J,K\le$5,6 in para-ammonia.  Both dipole ($\Delta K = 0$) and forbidden
($\Delta K\neq$0) transitions between the levels were included.

We  found that the inclusion of the $\nu_2$=1 rovibrational states has a
dramatic effect on the flux predicted for the observed pure rotational
transitions, when compared with predictions obtained without the inclusion
of vibrationally excited states.  However, the additional inclusion of the
$\nu_1$=1, $\nu_2$=2,$\nu_3$=1, and $\nu_4$=1 vibrational states has a
negligible effect on the computed fluxes for the observed transitions, a
maximum increase of only 2 percent being obtained for the flux of the ground
rotational transition of ortho-NH$_{3}$ with even smaller increases for the
remaining emission lines.  This conclusion, which is based on a model of
a molecular structure limited to rotational levels up to $J=7$ in each
vibrational state, is explained by the fact that the vibrational transition
moments to the symmetric bending state are significantly higher than those
to the remaining vibrational states.  In fact, the radiative pumping rate in
the $\nu_2$ mode dominates over radiative pumping rates in the remaining
modes, even in the innermost parts of the envelope where shorter-wavelength
radiation prevails.  At larger distances from the star, dust opacity
decreases the photon density at shorter wavelengths more effectively,
further reducing the relative effect of radiative pumping in other modes. 
The IR pumping effects for ammonia have been investigated previously by
\cite{Schoier2011} and \cite{Danilovich2014}.

Solving the radiative transfer, we do not distinguish between photons
produced in the envelope and photons coming from the central star.  However,
by comparing  the radiative rates from our best fit model with those 
obtained when only stellar photons are present, we were able to estimate
their relative importance.  Only in the inner part of the envelope is radiative
excitation in the $\nu_2$ transition dominated by the stellar photons. 
Very quickly, already at seven stellar radii the contribution from the
envelope begins to prevail over the contribution from the central star,
reaching a ratio of twenty at the outer edge.

Our final model for ammonia includes all levels up to $J=15$ in both the
vibrational ground state and the $\nu_2$=1 vibrational state, amounting to a
total of 172 levels and 1579 transitions for ortho-NH$_3$ and 340 levels and
4434 transitions for para-NH$_3$.  We include the ground state and
vibrational levels up to 3300 cm$^{-1}$, corresponding to 4750\,K, above
ground.  The completeness of the levels used in the computations may be
judged by comparing the partition function calculated for the model with
that obtained 
from
the full list of lines from the BYTe computations
\citep{Yurchenko2011}.  With the number of levels given above, the partition
function for the ortho species is equal to the BYTe number for temperatures
to 300~K, and is lower by $2\%$, $28\%,$ and $40\%$, respectively, at 600,
1000, and 1200~K.  The partition function of para species is also complete
up to 300~K, and is lower by $10\%$, $40\%,$ and $55\%$, respectively, at
600, 1000, and 1200~K.  The incompleteness in the levels may result in an
overestimate of their populations in the dense and hot inner parts of
the envelope.

We adopted the collisional rate coefficients from \cite{Danby1988}.  
The rate coefficients for collisional de-excitation
are available only for low-lying levels below $J=6$ and for
a maximum temperature of 300 K.  Collisional rates were extrapolated to 
higher temperatures with a scaling proportional to the square root of the gas
temperature. The extension of collisional rates to other levels is necessarily quite
crude.  At first we neglected collisional rates between levels not
available in Danby's work.  In this case, the ladder of levels with $K$ higher
than 6 in ortho-ammonia and higher than 5 in para-ammonia are populated only
by forbidden radiative transitions.  To analyse the influence of the unknown
collisional rates on the excitation of NH$_3$, we carried out additional computations 
based on crude estimates of the collisional rates; here, we adopted collisional
depopulation rates of 10$^{-11}$ or 10$^{-10}$\,cm$^{3}$\,s$^{-1}$ for rotational states
in the ground vibrational state, and rates of 
$10^{-14}$\,cm$^{3}$\,s$^{-1}$ for excited vibrational states. 
None of these approximations seem to
significantly influence our conclusions inferred from the modelling of the
observed transitions.

\section{Modelling}

Here, we present our numerical code and then describe all the assumptions made and
parameters investigated during our search for the best fits to the observed
rotational transitions of ammonia.  After that, we describe the method used to
search for the best fit to the data, and present the results thereby obtained.

\subsection{Numerical code and procedures}

To model circumstellar absorption and emission lines, we  developed a
numerical code, MOLEXCSE (Molecular Line EXcitation in CircumStelar
Envelopes), to solve the non-LTE radiative transfer of molecular lines and
the dust continuum.  Here, the methodology was to consistently include the effects
of optical pumping by the central star and infrared pumping by
circumstellar dust upon the population of the molecular levels.  The first
effect is more critical in the case of post-AGB stars, and the second one
more important in AGB stars \citep[see e.g.][]{TruongBach1987}.  The
effect of the dust on the radiation intensity inside the envelope is
included by solving the radiative transfer for the line and
continuum radiation.  For this purpose, the critical properties of the dust -- 
the coefficients of
extinction, scattering, and thermal emission -- are determined by a separate
code by modelling the observed spectral energy distribution of the source, 
as described in Section 4.4.  Reproducing the 
observed continuum fluxes allows the absorption 
line profiles for mid-infrared vibrational transitions to be computed.

The radiative transfer equation is formulated in the comoving frame
\citep{Mihalas1975} to include the effects of an expanding spherical
shell.  The only difference is the linearization of the differential equations
along tangent rays on a geometrical grid instead of on 
grid of optical depths.  
The purpose of this modification was to avoid numerical problems when
maser or laser transitions emerge during the iterations.

The simultaneous solution of the statistical equilibrium equations and of
the radiative transfer in lines is a non-linear process and, to be efficient,
requires linearization of the equations.  Full linearization of the
radiative transfer equation is very complex and its solution is
time-consuming.  In the code we follow the approach presented by
\citet{Schoenberg1986}.  The original formulation of their approximate
Newton-Raphson operator was modified to include the geometrical formulation
of radiative transfer mentioned above.

The program possesses two additional features useful in the application to
ammonia.  First, it enables the simultaneous solution of the radiative transfer
problem for more than one molecule.  Second, the radiative transfer may be
solved with the inclusion of line overlap effects between different
molecules.  With these features, one can consistently compute the effects of line
blending between ortho- and para-NH$_3$.  The prominent example of such a case
is the overlap between rotational transitions of para-NH${_3}$
3$_1$(s)-2$_1$(a) and of ortho-NH${_3}$ 3$_0$(s)-2$_0$(a).

In computing the profiles of the inversion lines, we resolved the hyperfine
structure of ammonia, including the effects of quadrupole and magnetic
splitting, as measured by \citet{Kukolich1967} and updated by
\citet{Rydbeck1977}.  Line strengths were computed when necessary using the
formula given by \citet{Thaddeus1964}.  These theoretical line strengths are
in good agreement with the observed intensities both in laboratory
experiments \citep{Kukolich1967} and in observations of molecular clouds
\citep{Rydbeck1977}.  The magnetic splitting is much smaller than the
quadrupole splitting and has little effect on the line profiles and
therefore was neglected in our computations.  When solving the radiative
transfer, the populations of the sublevels are distributed according to the
statistical weights, i.e.  assuming local thermodynamical equilibrium (LTE)
between sublevels.  The relative LTE ratios of intensities as defined by
\cite{Osorio2009} were applied including effects of line overlap.  The code
has been tested by comparison with previously published models of AGB
circumstellar envelopes.  A detailed description of our code and the tests
we  conducted will be published elsewhere.

\subsection{Thermal and density structure of the envelope}

The detailed structure of IRC$+$10216's envelope has been the subject of many studies. 
We mention here only those based on the analysis of CO emission lines 
\citep{Crosas1997, Schoier:2002nx, Agundez2012, deBeck2012} or the spectral energy
distribution \citep[e.g.][]{Menshchikov:2001kl}.

\begin{figure}
\centering
\includegraphics[width=8.5cm]{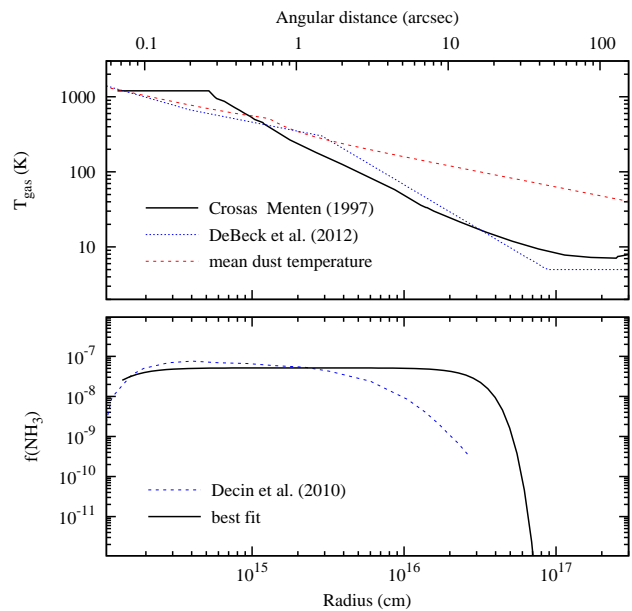}
\caption{
Gas temperature within the envelope of IRC$+$10216 adopted for this work
(upper panel) from \cite{Crosas1997} is shown by a solid line, while that
from \cite{deBeck2012} is shown by a blue dotted line.
The dust temperature is shown by a red dashed line.
The lower panel shows the
distribution of ammonia from our best fitting models (solid line),
and its theoretical distribution (dashed line) from \cite{Decin:2010kl}. 
The angular distance is given in the top axis of the upper panel for an assumed
distance to IRC$+$10216 of 130 pc.
}
\label{figmod}
\end{figure}

Several models for the thermal structure have been proposed to explain the
CO emissions from IRC$+$10216.  These models differ in the assumed distance,
mass loss rate and variation of gas velocity in the inner parts of the
envelope.  For the purpose of this paper, we have chosen, for the basic
temperature structure of the envelope, the model of \cite{Crosas1997}, but
with the distance reduced from 150 to 130 pc following
\cite{Groenewegen:2012fu} \citep[see also][]{Menten2012}.  The gas
temperature structure of \cite{Crosas1997} is based on self-consistent
computations of the temperature structure and radiative transfer in CO,
constrained by observations of CO emissions up to $J=6-5$.
 The adopted  distance, expansion velocity and mass loss rate are listed in Table\,2
and the gas temperature structure, $T_{\rm gas}$, is
presented in the upper panel of Figure~\ref{figmod} by the solid line. 
The temperature profile has been extended to the inner part of the envelope by
assuming a constant value of 1200~K.  
More recently, \cite{deBeck2012} used the GASTRoNOoM model (Decin et al.\ 2006; 2010) 
to derive the gas temperature structure shown by the dashed line 
in the upper panel of Figure~\ref{figmod}, and found it capable 
of explaining HIFI observations of CO up to $J=16-15$.
Figure~\ref{figmod} shows these
representative $T_{\rm gas}$ models to have distinct
temperature structures, which differ mainly in the middle part of the
envelope.

The density structure of the envelope is computed from the equation of mass
conservation, assuming that the gas is entirely composed of molecular
hydrogen and that the mass loss rate and outflow velocity are constant.  The
microturbulent velocity was set to be constant throughout the envelope and
equal to 1.0\,km\,s$^{-1}$ \citep{Crosas1997}.  In fact this value seems to
explain  the observed shape of the lowest ortho-NH$_{3}$
1$_0$(s)-0$_0$(a) emission well.

\subsection{Distribution of ammonia}

The distribution of ammonia in the circumstellar envelope is governed by the
poorly-understood process of ammonia formation, and the well-understood
process of its photodissociation in the outer part of the envelope.  A
detailed chemical model predicting the formation of ammonia and water as a
result of photo-processes in IRC$+$10216 was proposed by \cite{Agundez2010}
(see also \cite{Decin:2010kl}).  This model predicts peak abundances
of ammonia within the observed values but, as we show below, in its present
version cannot explain the observed intensities of the rotational
transitions.

For the purpose of this paper, we assumed that 
in the middle parts of the
envelope the maximum abundance of ammonia relative to H$_2$, $f_0$, is
constant, while in the inner and outer layers it is increasing and
decreasing, respectively.  To distinguish between ortho- and para-ammonia, 
their individual abundances are further designated $f$(ortho-NH$_3$) and 
$f$(para-NH$_3$).  The abundance profile adopted for 
the ammonia in the inner layers is based on the model of 
\cite{Decin:2010kl} (see their 
supplementary material).  
For each kind
of ammonia, we parametrized the increase of its abundance in the inner parts
of the envelope by two free parameters: the formation radius, $R_{\rm f}$, and
$f_{0}$.  To describe the increase in the
ammonia abundance in the inner envelope, we used a fit to the predictions of 
\cite{Decin:2010kl} of the form 
$f(r) = f_{0} \times 10^{-0.434 (R_{{\rm f}}/r)^{3}}$, where $r$ is
radial distance from the central star.  This parametrization, despite its
theoretical origin, is a more realistic description of the ammonia
distribution than a rather unphysical sharp increase of the ammonia abundance at
a fixed radius.  We note that $R_{\rm f}$ may be formally lower than the inner dust
shell radius.  To describe the decrease in the ammonia abundance near 
the photodissociation radius, $R_{\rm ph}$, in the 
outer layers of the
envelope, we used the analytical parametrization $f(r) = f_0 \times {\rm exp}(-{\rm ln} 2
(r/R_{\rm ph}$)$^\alpha$), which is frequently used in a parametrization of
CO photodissociation \citep[see e.g.][]{Schoier2001}.  A slope $\alpha =
3.2$ was adopted here, in accord with the fit to a detailed chemical
model of NH$_{3}$ photodissociation computed using the CSENV code by
\cite{Mamon1988}.  The ortho- and para-ammonia abundances, the inner radius
of ammonia formation, $R_{{\rm f}}$, and its photodissociation radius,
$R_{{\rm ph}}$, are free parameters in the fitting procedure.
The distribution of ammonia from our best-fit models and from
\cite{Decin:2010kl} are shown in Fig.~\ref{figmod} by the solid and dashed
lines, respectively.

\subsection{Dust shell model of IRC$+$10216}

\begin{figure}
\centering
\includegraphics[width=8.5cm]{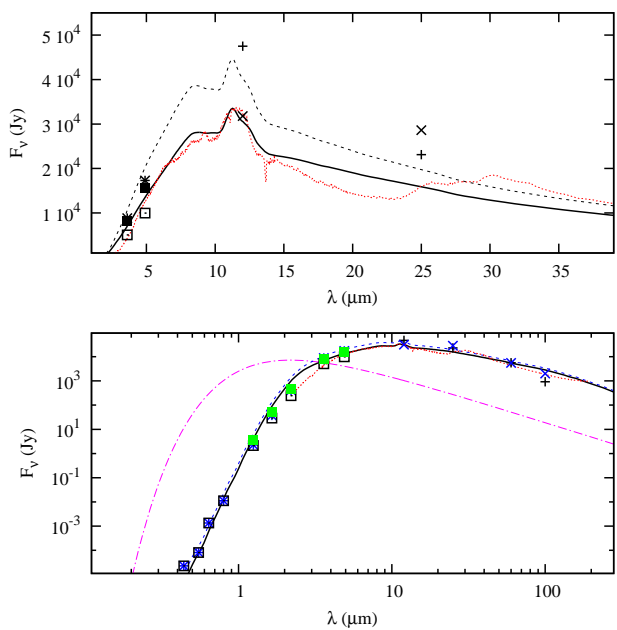}
\caption{
Fit to the flux for L$_{\rm avg}$=8500\,L$_\odot$ and 
effective temperature of 2300 K is shown in both panels by black solid line. 
The contribution from the central star is shown in the bottom panel 
by the dash-dotted magenta line.
SWS and LWS ISO spectra are shown by red dotted lines. Photometric data 
below  1\,$\mu$m are from GSC 2.3, those between 1.25 and 5 $\mu$m are 
from \cite{LeBertre1992} at three different phases: 
0.02 (stars), 0.21 (filled squares) and 0.23 (dotted squares), 
while at longer wavelengths the flux is from IRAS
(crosses) and COBE DIRBE (x) measurements. 
The dust model obtained for L$_{max}$=11850\,L$_\odot$
is shown by  dashed lines on each panel  (see text for details).
}
\label{figd}
\end{figure}

As  explained in Sect.\,3, the excitation of ammonia is dominated by
radiative pumping via the 10\,$\mu$m $\nu_2$=1 band.  Therefore, to model
the ammonia lines we need to estimate the continuum flux at
10\,$\mu$m inside the whole envelope.  For that purpose, we used a dust radiative
transfer model \citep{Szczerba:1997eu}, which is able to provide all
necessary physical variables to MOLEXCSE.  We fit a combination of flux
measurements from the Guide Star Catalogue ver.\ 2.3 (GSC 2.3) for
$\lambda\,<\,1\mu$m, photometric data between 1.25 and 5 $\mu$m from
\cite{LeBertre1992}, and at longer wavelengths from IRAS and COBE DIRBE.  In
addition, as the strongest constraint, we have used ISO Short Wavelength
Spectrograph (SWS) and Long Wavelength Spectrograph (LWS) spectra.  They
were obtained on May 31 1996 (JD=2450235) and correspond to phase $\phi$ of
0.24, counted from the reference maximum phase of the light curve, $\phi$=0,
on November 17 1988 (JD=2447483), and given a period of 649 days
\citep{Menshchikov:2001kl}.  We have used this older parameterization of the
IRC$+$10216 variability as it was obtained closer to the date of
the ISO observations.  For this phase of pulsation and the distance of 130 pc
determined recently by \cite{Groenewegen:2012fu}, the average luminosity of
IRC$+$10216 was estimated to be about 8500 L$_\sun$
\citep{Menshchikov:2001kl, Menten2012}.

As dust constituents, we adopted amorphous carbon of AC type from
\cite{Rouleau:1991nx} and SiC from \cite{Pegourie:1988eu}, both with a
power-law distribution of grain radii with an index of $-$3.5 between 0.1
and 0.43\,$\mu$m.  We assumed a maximum allowed dust temperature of 1300\,K. 
We assumed a constant outflow velocity equal to the observed terminal
velocity of 14.5\,km\,s$^{-1}$, in spite of a clear variation of the
outflowing velocity in the inner part of envelope seen in higher molecular
transitions \citep{Agundez2012}, and a constant dust mass loss rate.  The
derived dust mass loss rates were $1.0\times10^{-7}$ M$_\odot$ yr$^{-1}$ and
$3.0\times10^{-9}$ M$_\odot$ yr$^{-1}$ for AC and SiC dust, respectively. 
The inner shell radius corresponding to the assumed maximum dust temperature
is reached at a distance of about 3 stellar radii.  The required total
optical depth of the envelope at V is 18.5 magnitudes, which corresponds to
$\tau_{10{\mu}m}$ equal to 0.18.  The parameters of our model for ammonia in
IRC$+$10216 are listed in Table~\ref{TableModel} and the mean dust
temperature is shown by red dashed 
line
in Figure~\ref{figmod}.  As is shown
below, the dust temperature determines the radiation intensity in the
continuum, which is important for vibrational pumping.

\begin{table}
\caption{Model for IRC$+$10216}
\label{TableModel}
\begin{tabular}{ll}
\hline\hline
Parameter & Value \\
\hline
Distance       & $d=130$ pc \\
Expansion velocity & $v_{\rm exp}=14.5$ km s$^{-1}$ \\
Mass loss rate (total hydrogen) & ${\dot{M}}=3.25\times10^{-5}$ M$_\odot$ yr$^{-1}$\\
Stellar radius & $R_{\star} = 3.9\times10^{13}$ cm \\
Effective temperature & $T_{\rm eff}=2300$ K \\
Luminosity (average)     & $L_{\rm avg} = 8500$ L$_{\odot}$ \\
Luminosity at maximum & $L_{\rm max} = 11850$ L$_{\odot}$\\
Inner shell radius  &  $1.2\times10^{14 }$ cm ($\sim$3\,$R_{\star}$)\\
\,
{\it Best-fit parameters:} & \\
Formation radius $R_{\rm f}$  &  $1.0\times10^{14 }$ cm 
~~($\sim$2.5\,R$_{\star}$)\\
Photodissociation radius $R_{\rm ph}$   & $> 2-3\times10^{16}$ cm \\
$f$(ortho-NH$_3$) & $(2.8\pm0.5)\times10^{-8} (\frac{3.25\times10^{-5}}{\dot{M}})$ \\
$f$(para-NH$_3$) & $(3.2^{+0.7}_{-0.6})\times10^{-8} (\frac{3.25\times10^{-5}}{\dot{M}})$ \\
\hline
\hline
\end{tabular}
\end{table}

The fit obtained for a stellar luminosity $L_{\rm avg}$=8500\,L$_\odot$ and
an effective temperature of 2 300 K is shown in both panels of Fig.\ref{figd}
by the black solid line.  In our modelling, we did not include MgS, which is
commonly used for modelling of the 30\,$\mu$m structure, since this material has
no optical constants in the optical range so the precise determination of
its temperature is impossible \citep[see e.g.][]{Szczerba:1997eu}.  However,
since we are interested in the continuum emission at 10\,$\mu$m, this
approach seems to be justified.

Most of the ammonia lines were observed at phase $\phi=0.23$ (see Table\,1),
when the luminosity of the central star was close to its average value
$L_{\rm avg}$=8500\,L$_\odot$.  However, three of the lines were observed at
$\phi \sim 0.13$ so, for the purpose of ammonia line modelling, we rescaled
their observed integrated area down to $\phi$=0.13 using the predicted
variability from models obtained at maximum and at average luminosity. 
Since we do not have ISO spectra taken at $\phi$=0, we cannot constrain the
dust properties at the maximum stellar luminosity.  Therefore, we 
decided to recompute our dust model by keeping all parameters (including the
inner radius of dust shell) constant, except for the stellar luminosity,
which was raised to $L_{\rm max}$.  The model results thereby obtained are
shown by dashed lines in both panels of Fig.\,\ref{figd}.

After modelling the dusty envelope, we exported the dust thermal emission
coefficient, along with the total extinction and scattering coefficients, to
the code MOLEXCSE, as a function of wavelength and radial distance. 
MOLEXCSE, which was described above in Sect.\,4.1, was then used to solve
the multilevel radiative transfer in lines and continuum.  See Eq.\,(1) in
\cite{Szczerba:1997eu} for details concerning the exported physical
quantities.

\subsection{The best-fit models}

Using our code described in Sect\,4.1, the ammonia abundance profile given
in Sect.\,4.3, the model parameters specified in Table\,2, together with the
gas temperature distribution from \cite{Crosas1997} and, taking into account
vibrational pumping by infrared radiation, we computed a grid of separate
models for ortho-NH$_3$ and para-NH$_3$.  The grid of models was computed
for formation radii $R_{\rm f}=2.5, 5, 10, 20, 40,$ and 80 R$\star$,
photodissociation radii $R_{\rm ph}$ ranging from $1.5\times10^{16}$\,cm to
$6\times10^{16}$\,cm in steps of $\Delta R_{\rm ph}=0.5\times10^{16}$\,cm,
and $f$(ortho-NH$_3$) and $f$(para-NH$_3$) ranging from $1.4\,10^{-8}$ to
$4.8\,10^{-8}$ in steps of $\Delta f$(NH$_3$)$=0.2\,10^{-8}$.

For each model we defined the figure-of-merit as \begin{equation} \chi^2 =
\sum_{i=1,N_{\rm lines}} \left(\frac{F_{i}^{\rm obs}-F_{i}^{\rm
th}}{\sigma_{i}}\right)^{2}, \end{equation} where $F_i^{\rm obs}$ (see
Table\,1) and $F_i^{\rm th}$ are the observed and theoretically predicted
integrated fluxes for the $i$-th line, respectively, $\sigma_i$ (see
Table\,1) is the estimated uncertainty in $F_i^{\rm obs}$, and N$_{\rm
lines}$ is 3 for ortho- and 6 for para-NH$_3$ (see Fig.\,2).  Here, the
three lines observed in 2011 at phase $\phi$=0.13 were rescaled to phase
$\phi$=0.23, assuming that they vary co-sinusoidally between the maximum and
average luminosity of the central star.  From the computed integrated line
fluxes for these stellar luminosities, we obtained the amplitude of their
variations as 4.0, 3.1, and 8.2 K\,km\,s$^{-1}$ for the transitions at
1763.524, 1763.601, and 1763.823 GHz, respectively.  Since the value of the
cosine function describing $\phi$ decreases by about 0.6 between phase 0.13
and 0.23, we reduced the integrated fluxes given in Table\,1 by 0.6 times
the estimated amplitude of the line variations, i.e.  by 2.4, 1.9, and 5.0
K\,km\,s$^{-1}$.  The quality of each fit is measured by the reduced
$\chi^2$ parameter, which for two free parameters is defined here as
$\chi^2_{red} = \chi^2 / (N_{\rm lines}-2)$.

\begin{figure*}[]
\centering
%
\includegraphics[width=5.8cm]{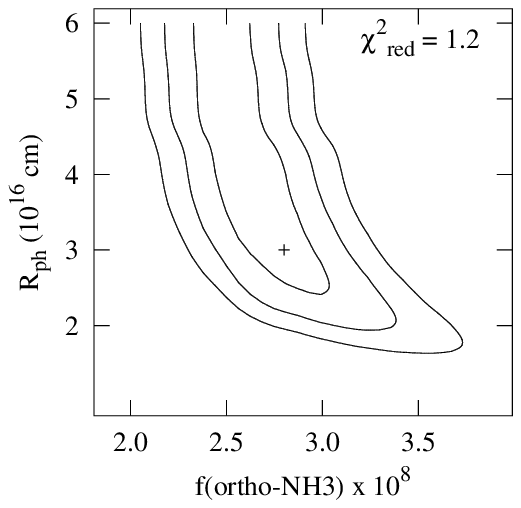}
%
\includegraphics[width=5.8cm]{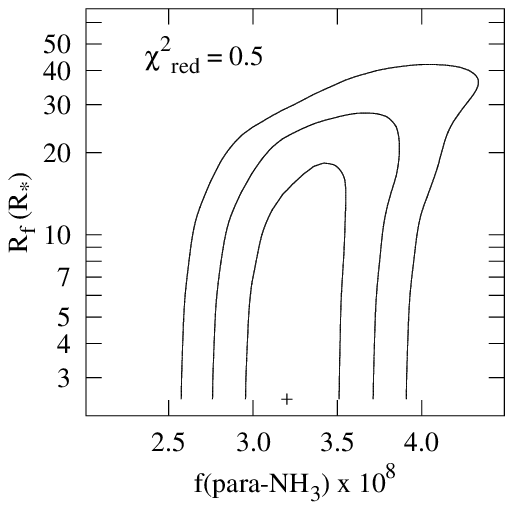}
%
\includegraphics[width=5.8cm]{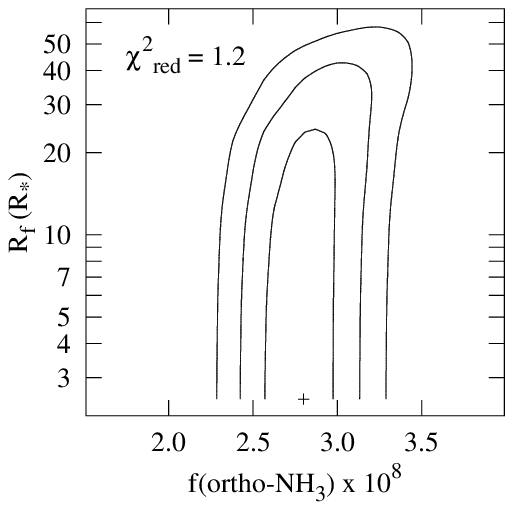}
\caption{
Contours of $\chi^2$, indicating the sensitivity of the fit to the
model's free parameters.  The left panel shows the dependence of $\chi^2$ on the
photodissociation radius $R_{\rm ph}$ and abundance of ortho-NH$_3$, given the 
best fitting value for the formation radius, $R_{\rm f}$ = 2.5 $R_\star$.  
The middle and the right
panels show the dependence of $\chi^2$ on the formation radius $R_{\rm
f}$ and the abundances of para- and ortho-NH$_3$, given a photodissociation
radius $R_{\rm ph}$ = 3\,10$^{16}$\,cm.  The contours correspond to 1, 2, and 3 $\sigma$
confidence levels.  The best fits are marked with a cross and the
corresponding $\chi^2_{red}$ are shown on the plots.
}
\label{figchisquare}
\end{figure*}

We  found that for $R_{\rm ph}\geq$\,3$\times$10$^{16}$\,cm, the minimum
value of $\chi^2_{\rm red}$ is achieved for $R_{\rm
f}=2.5$\,R$\star=1.0\times10^{14}$ cm, which is the minimum formation radius
possible for the assumed inner radius of the envelope considered during
modelling.  However, since the assumed $R_{\rm ph}$ is decreased, the formation radius
that minimizes $\chi^2_{\rm red}$ increases (from $R_{\rm
f}=10$\,R$\star$ for $R_{\rm ph}=$\,2.5$\times$10$^{16}$\,cm to $R_{\rm
f}=40$\,R$\star$ for $R_{\rm ph}=$\,1.5$\times$10$^{16}$\,cm); at the same time, however, 
the minimum $\chi^2_{\rm red}$ also increases.  Hence, the value of the
photodissociation radius was found by searching for the minimum $\chi^2_{\rm
red}$ among the models for $R_{\rm f}=2.5$\,R$\star$.

The best fit to the ortho-NH$_3$ lines is achieved for $R_{\rm ph} >
2-3\times10^{16}$\,cm, and $f$(ortho-NH$_3)=(2.8\pm0.5)\times10^{-8}$ at a
3\,$\sigma$ confidence level.  There is no strong constraint on the maximum
value that $R_{\rm ph}$ could have.  To estimate the errors, we constructed
a $\chi^2$ map for ortho-NH$_3$, which is presented in the left panel of
Fig.\,5, and shows how $\chi^2$ varies as a function of the
photodissociation radius $R_{\rm ph}$ and $f$(ortho-NH$_3$) for the best-fit
formation radius $R_{\rm f}=1.0\times10^{14}$ cm.  The contours correspond
to 1, 2, and 3\,$\sigma$ confidence limits.  The best fit is indicated with the
cross and the corresponding value of $\chi^2_{\rm red}=1.2$ is shown on the
plot.  Rotational transitions of para-NH$_3$ are even less sensitive to the
photodissociation radius, and a plot similar to that shown on left panel of
Fig.5 does not yield any constraints on $R_{\rm ph}$.  This behaviour could
be related to the fact that at a given distance from the star the pumping
mid-IR radiation is the same, while at low gas temperatures the excitation
of the 1$_0$(s)-0$_0$(a) transition of ortho-ammonia is much more efficient
than the excitation of the 2$_1$(s)-1$_1$(a) transition of para-ammonia (see
Fig.1).  On the other hand, the lowest para-NH$_3$ transition,
$2_1$(s)-$1_1$(a), was observed with HPBW of 18.2\arcsec, corresponding to a
projected radius of 1.8$\times$10$^{16}$ cm at the assumed distance of 130
pc; thus, for values of R$_{\rm ph}$ that are larger than this projected
radius, this line is less sensitive to R$_{\rm ph}$ than  the
1$_0$(s)-0$_0$(a) ortho-NH$_3$ transition, for which the HPBW is twice as
large.

The derived abundance of para-NH$_3$ for the best-fit model is
($3.2^{+0.7}_{-0.6})\times10^{-8}$, which means that, to within the error
bars, the ratio of ortho- to para-NH$_3$ is equal to 1, a ratio that is
characteristic of the formation of NH$_3$ at high temperatures
\citep{Umemoto:1999bs}.  The average abundance for the two species is
(3.0$\pm0.6)\times 10^{-8}$.  The error estimate for the para-NH$_3$
abundance given above is the 3\,$\sigma$ confidence interval obtained from
the $\chi^2$ map for para-NH$_3$, which is presented in the middle panel of
Fig.\,5 and shows how $\chi^2$ varies as a function of the formation radius
$R_{\rm f}$ and abundance of para-NH$_3$ at the best-fitting
photodissociation radius $R_{\rm ph}$ = $3\times10^{16}$\,cm.  The meaning
of the contours is the same as in the left panel.  The best fit, indicated with
the cross, and the corresponding $\chi^2_{\rm red}$ are also shown on the
plot.  The minimum $\chi^2_{\rm red}$ of 0.5 implies an excellent fit to the
six para-NH$_3$ lines (with four degrees of freedom).

While the $\chi^2$ analysis described above made use of integrated line
fluxes instead of the detailed line profiles, the model successfully reproduces the
line shapes.  The emission line profiles resulting from the best-fit
models are shown in Fig.\,2 by dashed lines, and the best-fit parameters
are listed in Table\,2.  As indicated in Table 2, the abundances derived for 
ortho- and para-NH$_3$ are inversely proportional to the assumed mass-loss rate;
this can be understood as a consequence of the facts that the NH$_3$ lines are 
optically thin and that the derived ammonia abundance  is only weakly dependent 
on the exact choice of the gas temperature (see below), so the same total mass 
of ammonia is required to explain the observations regardless of the mass-loss rate.

To check how strongly the derived best-fit parameters depend on the assumed
gas temperature inside the envelope, we performed some test computations
using the $T_{\rm gas}$ structure from \cite{deBeck2012}.  For the best
fitting parameters found above and the $T_{\rm gas}$ structure of
\cite{deBeck2012}, we found that the line shapes are only slightly
different, but $\chi^2_{\rm red}$ increases to 3.8 for ortho- and to 4.6 for
para-NH$_3$.  Searching for the model minimizing $\chi^2_{\rm red}$, we found
that the best fit is obtained for a formation radius $R_{\rm
f}$=$1\times10^{14}$\,cm, a photodissociation radius $R_{\rm ph}$ =
$3\times10^{16}$\,cm, and abundances of ortho-NH$_3$ and para-NH$_3$ equal to
($3.0^{+0.6}_{-0.5})\times10^{-8}$ and ($3.6^{+0.8}_{-0.7})\times10^{-8}$,
respectively.  These values lie within the error bounds estimated 
using the \cite{Crosas1997} temperature profile,  implying that the abundances 
and photodissociation radius derived from rotational transitions are only weakly dependent
on the choice of temperature profile.

Although we obtained the best-fit results based solely on rotational
transitions, our model also offers predictions for the other observed
transitions.  Therefore to make a quality test of our best-fit models,
we examined whether they can reproduce both the inversion and mid-IR transitions of
ammonia.

\begin{figure}
\centering
\includegraphics[width=8.5cm]{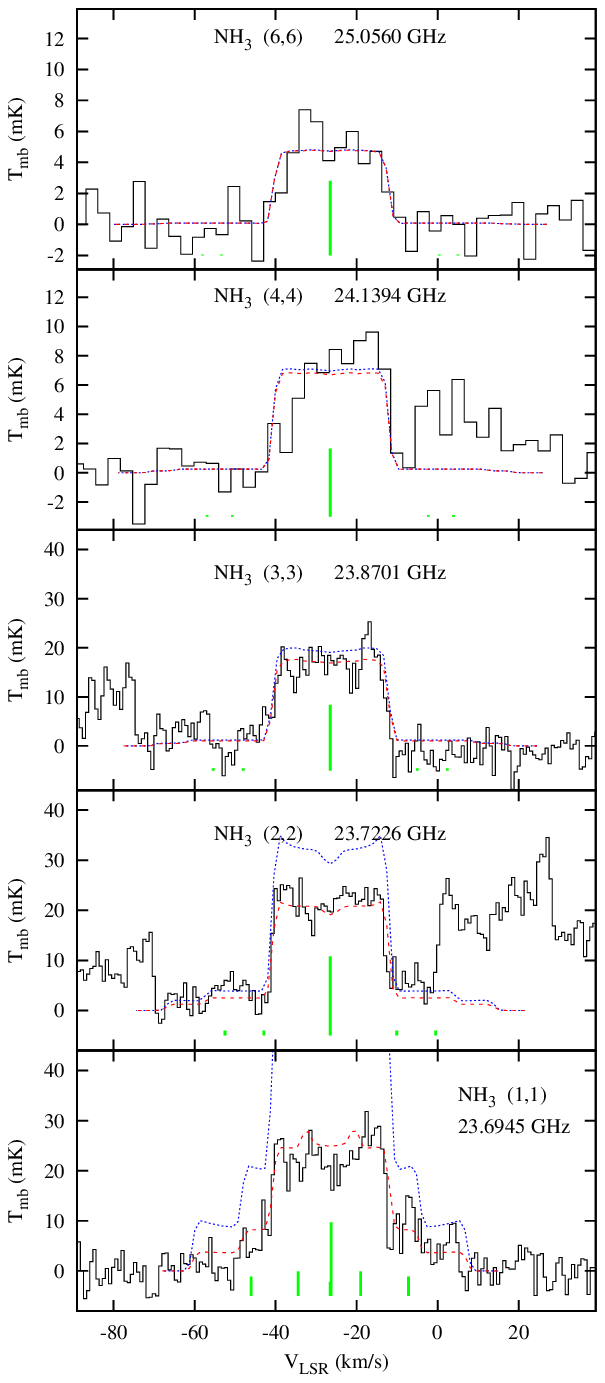}
\caption{
Profiles of the inversion lines as computed for the best-fit models, which
assume $R_{\rm ph} = 3\times10^{16}$ cm and the $T_{\rm gas}$ distribution from
\cite{Crosas1997}, (blue dotted lines) are overplotted on the observed inversion
transitions (solid lines) from \cite{Gong2015}.  
Theoretical profiles for the same $T_{\rm gas}$ profile, but
with the photodissociation radius reduced to $R_{\rm ph} = 1.5\times10^{16}$\,cm
are shown with red dashed lines. Vertical sticks show positions and relative
intensities of hyperfine components.
}
\label{figinv}
\end{figure}

While \cite{Gong2015} have obtained observations of the inversion transitions
at several different phases (in range 0.20-0.96), our models do not predict any dependence
on the integrated fluxes 
in
the phase (i.e. luminosity) of the central star.  
Calculations show that the expected variation of flux in the (1,1)
inversion line is up to 3 percent between
phases of maximum ($\phi$=0) and mean ($\phi$=0.25), and is even lower in higher inversion
transitions.
Thus, our model computed for the average luminosity of the star
can be used for comparison with
the inversion line observations.  Figure\,\ref{figinv} shows the observed transitions by
means of solid lines \citep{Gong2015} and the computed theoretical profiles
with dashed lines.
Vertical sticks mark positions of hyperfine components - the height of
the central stick is chosen arbitrary, while the heights of the other sticks
are scaled according to their relative theoretical intensities.
Our best-fit model for the rotational transitions fits
the upper inversion transitions very well, but overestimates the flux
measured for the para-NH$_3$ (2,2) line, and significantly overestimates
that for the lowest para-NH$_3$ (1,1) line.  We have found that
decreasing $R_{\mathrm{ph}}$ to 1.5$\times10^{16}$\,cm gives a very good fit
to the all observed inversion lines (shown by dotted lines in
Fig\,\ref{figinv}).  However, this model is not able to match
the rotational lines as successfully as our best-fit model.  
Using the best-fit abundance of ortho-NH$_{3}$ and decreasing the 
photodissociation radius to 1.5$\times10^{16}$\,cm 
significantly increases the $\chi^2_{\rm red}$, to a  value 28.  
On the other hand, the best-fit model with fixed R$_{\rm
ph}$=1.5$\times10^{16}$\,cm and free formation radius and abundance of
ortho-NH$_{3}$ moves R$_f$ to about 40 R$_\star$ and
f(ortho-NH$_3$) to 4.2$\times$10$^{-8}$ with $\chi^2_{\rm red}$ equal to
3.8.

To investigate this discrepancy, we  extended the search for the best-fit model by adding the inversion lines.  To this purpose, we  took the
integrated flux from \cite{Gong2015} and converted it to K\,km\,s$^{-1}$ scale. 
However, our theoretical fit to the inversion lines seems to suggest the
presence of hyperfine components, as we describe below in Section 5.4 of
discussion, at least in the case of the two lowest inversion lines, (1,1)
and (2,2).  Therefore, we  repeated the calculation of the integrated
flux of these two lines, extending the integration range accordingly.  This
increased the integrated flux of those two transitions up to about 20\%\ in case the of (1,1).  The integrated fluxes used for the analysis
of the inversion lines for ortho-NH$_3$ (3,3) and (6,6) transitions are
0.$47\pm0.5$ and $0.15\pm0.03$ K\,km\,s$^{-1}$, respectively, while for the
para-NH$_3$ (1,1), (2,2), and (4,4) transitions are $0.84\pm0.8$,
$0.64\pm0.9$, and $0.20\pm0.4$ K\,km\,s$^{-1}$, respectively.  Now,
considering simultaneously the inversion and rotational lines of para-NH$_3$
put a strong constraint on the photodissociation radius.  For the formation
radius fixed to the $R_{\rm f}=1.0\times10^{14}$ cm, we  found that the
photodissociation radius is R$_{\rm ph}=(1.5\pm0.2)\times10^{16}$\,cm and
the abundance of para-NH$_{3}$ f(para-NH$_3$) = $(3.2\pm0.5)\times$10$^{-8}$
with $\chi^2_{\rm red}$ equal to 0.86.  However, the same approach for
ortho-NH$_3$ gives rather a different estimation of the photodissociation
radius, which is equal to $(5_{-3.}^{>+1.})\times10^{16}$\,cm and the
abundance of ortho-NH$_{3}$ equal to $(2.4\pm0.4)\times$10$^{-8}$ with
$\chi^2_{\rm red}$ equal to 1.9.  Thus, we see, that a global fit gives ortho-
and para-NH$_3$ abundances that are quite similar (within error bars) to those
obtained from the analysis limited to the rotational lines only.  However,
the problem of requiring different photodissociation radii for ortho- and
para-NH$_3$ remains.

\begin{figure}
\centering
\includegraphics[width=8.5cm]{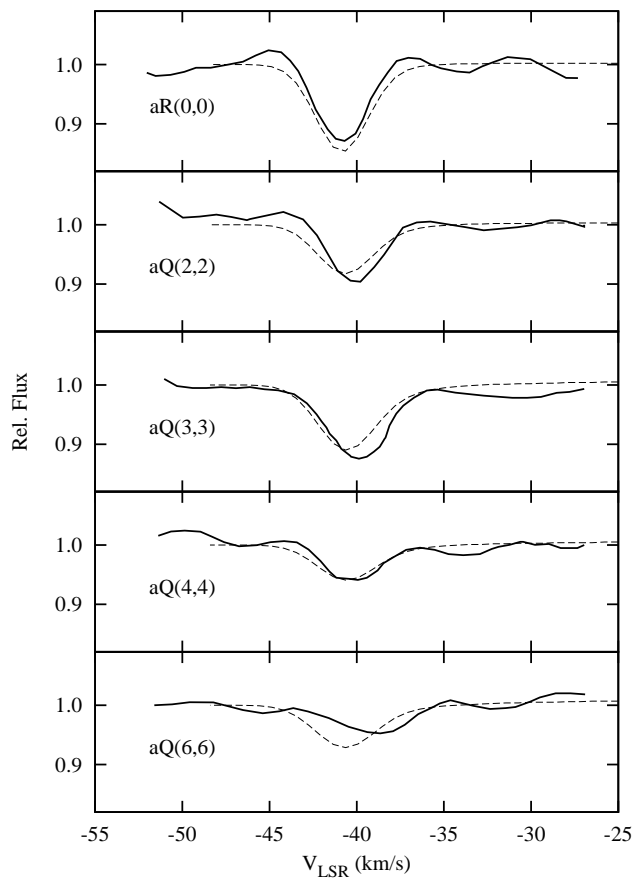}
\caption{
Profiles of NH$_3$ lines in the $\nu_2$ band observed by \cite{Keady1993}
(solid lines) and synthesized profiles using data from the best-fit models
(dashed lines).  Theoretical lines have been convolved with a Gaussian profile
to achieve a spectral resolution of 2.9\,km\,s$^{-1}$.
}
\label{figir} 
\end{figure}

The infrared transitions to the $\nu_2$ levels are seen in absorption
against the background dust continuum emission.  The observed depths of
lines depend on the telecope beam size.  In addition, the line spectrum is
smoothed by the instrumental profile.  The observed $\nu_2$ line profiles
published by \cite{Keady1993} and synthesized using our best-fit models are
shown in Figure~\ref{figir}.  There is reasonable consistency between our
model results (dashed line) with the observed profiles (solid line).  Some
disagreement with the aQ(6,6) line is most probably due to the neglect of
the inner velocity structure of the envelope or  to a higher value of the
formation radius.

\section{Discussion}

\subsection{Abundance of ammonia}

Over the past 30 years, the determination of the NH$_3$ abundance in the
envelope of IRC$+$10216 has been the subject of observational studies at
radio, mid-infrared, and submillimeter wavelengths.  All these observations
appeared to provide divergent abundances but, as we discuss below, the main
reason for the different results comes from the inclusion (or not) of
radiative pumping to the vibrational levels and differences in column
densities.

Prior to this study, the only observation of a NH$_3$ rotational transition
was made with the spectrograph on board the 
{\it Odin} satellite
\citep{Hasegawa2006}.  The rather poor signal-to-noise ratio obtained for the lowest
$1_0$(s)-0$_0$(a) transition of ortho-NH$_3$ did not provide a
precise measurement of the line profile.  The ammonia abundance relative to H$_2$
was determined as being $1\times 10^{-6}$, significantly above our
determination, $f$(ortho-NH$_3$)=$2.8\times10^{-8}$.  However, the analysis
of the line was based on a simplified model of a single $K=0$ ladder, limited
to the ground vibrational state and, more importantly, neglecting 
radiative pumping by the infrared radiation.  For the
adopted model of the IRC$+$10216 envelope, results from our code show that, with the
neglect of IR pumping
the abundance of ammonia has to be increased by factor of about 20 to
explain the measured 1$_0$(s)-0$_0$(a) line flux.  In addition, the
line profile becomes parabolic, which is characteristic of unresolved
optically thick emission, in disagreement with the profiles observed by {\it
Herschel}.  Thus, we can conclude that the largest discrepancy in the
determination of the ammonia abundance seems to be resolved.  
The effect of infrared pumping on the molecular abundances derived for
evolved stars was previously noted
in the paper of \cite{Agundez2006}, who show that by including the
$\nu_2$ mode of H$_2$O the abundance of water was decreased by a factor of $\ge 10$
with respect to that derived assuming pure rotational excitation.  For the specific case
of NH$_3$, \cite{Schoier2011} have shown that taking into account IR
pumping via vibrationally excited states will decrease the abundance of ammonia
by an order of magnitude.  In addition, \cite{Danilovich2014} have derived
the abundance of ammonia in the S-type star W Aql, including the $\nu_2$
state in their analysis of the radiative transfer.

The early observations of the (1,1) and (2,2) inversion transitions of
para-NH$_3$ around 23\,GHz were made by \cite{Kwok1981, Bell1982}, and
\cite{Nguyen-Q-Rieu1984}.  For their assumed distance of 200\,pc and mass
loss rate of $2\times10^{-5}$ M$_{\odot}$\,yr$^{-1}$, they found
$f$(para-NH$_3$) as being between $3\times10^{-8}$ and $2\times10^{-8}$.  At
the distance of 130\,pc adopted here, and for optically
thin transitions, the corresponding values of $f$(para-NH$_{3}$) are
$0.8\times10^{-8}$ and $0.5\times10^{-8}$, respectively, for the assumed
mass loss rate of $3.25\times10^{-5}$.  The estimates are a factor of 4
to 6 lower than our determination of para-NH$_{3}$ abundance, namely
$(3.2^{+0.7}_{-0.6})\times10^{-8}$.  This difference may be
due to simplifying assumptions made in the derivation of the ammonia abundance in
these earlier papers.  Recently, \cite{Gong2015} have inferred the abundances of
ortho- and para-NH$_{3}$ from observations of their metastable inversion
transitions $(J, K) = (1,1), (2,2), (3,3), (4,4), (6,6)$.  They lie between
$7.1\times10^{-8}$ to $1.4\times10^{-7}$ for ortho-NH$_3$, and
between $1.1\times10^{-7}$ and $1.3\times10^{-7}$ for para-NH$_3$.  Here, the
assumed mass-loss rate was $2\times10^{-5}$\,M$_{\odot}$ so, for our adopted mass
loss rate, the abundances correspond to values from $4.4\times 10^{-8}$ to
$8.6\times 10^{-8}$ for ortho-NH$_3$ and from $6.8\times 10^{-8}$ to $8.0\times
10^{-8}$ for para-NH$_3$.  Moreover, these values are dependent on the 
column density of molecular hydrogen, which is uncertain by a factor of 2
\citep{Gong2015}.

Observations of mid-IR $\nu_2$ transitions in absorption performed by
\cite{Betz1979, Keady1993} have yielded another estimate of the total abundance of
ammonia.  For an assumed mass-loss rate of $2\times10^{-5}$ M$_{\odot}$\,yr$^{-1}$, these
two studies inferred total ammonia abundances of $1\times10^{-7}$ and
$1.7\times10^{-7}$, respectively.  Rescaled to our adopted
mass-loss rate, these values transform to $6\times10^{-8}$ and $1\times10^{-7}$, 
in agreement with the total ammonia abundance derived by us 
$f$(NH$_3)=(6.0\pm0.6)\times10^{-8}$.

Finally, we note that the remaining differences -- 
at a factor of a few 
-- between the NH$_3$ abundances determined by various methods may not
be so significant.  The determination of the NH$_3$ abundances using mid-IR
absorption $\nu_2$ transitions requires an assumption for the column density
of ammonia.  The values range from about $2\times10^{15}$ cm$^{-2}$
\citep{Betz1979, Keady1993} to $8\times10^{15}$ cm$^{-2}$
\citep{Monnier2000b}.  This may be compared to the total column density
along the sight line to the central star in our best-fit model,
$1.5\times10^{16}$ cm$^{-2}$.  The different observations average the column
density over different parts of the envelope and therefore probe different
regions of the CSE.  We note that the column density is very sensitive to the
densest inner parts of the envelope.  For example, for our model with
constant velocity and with $R_{\rm f} = 10 R_{\star}$ the total column
density drops to only $5\times10^{15}$ cm$^{-2}$.  The assumed velocity
profile in the acceleration zone also plays a role.  For example,
\cite{Monnier2000b} had to assume a much higher column density of
$8\times10^{15}$ cm$^{-2}$, by modifying only the behaviour of the velocity
field in the inner part of the envelope, while using the model of the dusty
envelope similar to the original one from \cite{Keady1993}.

\subsection{Ammonia formation radius}

The best-fit models for ortho- and para-NH$_3$, marked by $+$ signs on the
middle and on the right panels of Fig.\,5, respectively, are obtained for
$R_{\rm f}=1.0\times10^{14}$ cm or 2.5~$R_{\star}$.  This means that a lower
limit on the ammonia formation radius is not provided by our model fits.  On
the other hand, the distribution of contours for 1, 2, and 3\,$\sigma$
levels on these panels show that ammonia cannot be formed too far from the
stellar photosphere.  At the 1$\sigma$ level, the upper limit on the
formation radius must be at most 20\,R$_{\star}$, while at the 2$\sigma$
level it is around 35\,R$_{\star}$.  This seems to be in rough agreement
with the formation radius of ammonia as estimated by \cite{Keady1993} and
\cite{Monnier2000b} on the basis of their analysis of mid IR lines.  The
analysis of \cite{Keady1993} allows a distance closer than 10\,R$_{\star}$,
whereas \cite{Monnier2000b} conclude that ammonia is formed at a distance
$\geq20$\,R$_{\star}$.  This last conclusion was based on interferometric
observations of the mid-IR bands in the continuum and at line centre and
further modelling of the ratio of visibility functions.

The rotational line profiles of NH$_3$ were fit assuming that the
velocity field is constant.  However, we performed test computations with
the gas velocity field inside the envelope used by \cite{Crosas1997}.  We
found that they produce a narrow emission feature in the centre of the
highest rotational and inversion transitions that are not seen in the
observed line profiles.  To remove this feature, created by the acceleration
region, it was necessary to constrain the formation radius by shifting
$R_{\rm f}$ above $5-10$ R$_{\star}$, although the quality of the fit -- as measured by
$\chi^2_{\rm red}$ -- was worse in this case.  Nevertheless, it seems that the
more realistic gas velocity field suggests that ammonia is not formed at the
stellar photosphere as  suggested by our formal fit to the
rotational transitions, assuming a constant outflow velocity.

\subsection{Photodissociation radius}

The photodisociation radius has been constrained by observations of the of
the NH$_3$ (1,1) and (2,2) inversion lines \citep{Nguyen-Q-Rieu1984}.  Their
analysis of the observed lines required a cut-off of the molecular abundance
beyond a radius of $(0.6-1)\times10^{17}$ cm at the assumed distance of
200\,pc.  This corresponds to $(4.0-6.5)\times10^{16}$\,cm at distance
assumed in this paper, d=130\,pc, and is in the range allowed by our
modelling, $R_{\mathrm ph}>2-3\times10^{16}$ cm.  Models for the NH$_3$
distribution in the circumstellar envelope of IRC$+$10216 presented by
\cite{Decin:2010kl}, and shown by the dotted line in Fig.\,3, predict a
decrease in the ammonia abundance in the outer envelope that is more rapid
than is allowed by the observations.  With this type of distribution, we cannot
fit the rotational lines of ammonia, especially the lowest transition of
ortho-NH$_3$, since its upper level is underpopulated at still quite high
gas temperatures (see Figs.\,1 and 3).  Similarly, this is also the case for
the reduced photodissociation radius inferred from the fit to the
observations of inversion lines by \cite{Gong2015}.

\subsection{Line shapes}

Thanks to the high signal-to-noise ratio (S/N)  obtained for the ortho-NH$_3$
$1_0$(s)-0$_0$(a) spectrum, the details of the line profile are
well-constrained (see bottom left
panel in Fig.\,2).  Another line, the $3_2$(s)$-2_2$(a) line at 1763.823 GHz,
was also observed at a high S/N ratio and has an identical shape (see the middle
right panel in Fig.\,2).  The energy of the upper level for these two
transitions is quite different (29 and 127\,K, see Table\,\ref{TableObs}),
and the similarity of their shapes suggests that both of them are excited in
a rather similar region within the envelope.  Surprisingly, their shapes show two
peaks resembling the profiles of resolved optically thin transitions. 
However, this is not possible because, given the {\it Herschel} beam size at the
frequency of the $1_0$(s)-0$_0$(a) transition, it would require an outer
radius for the ammonia-emitting region to be in excess of $10^{17}$ cm, a value close to
the CO photodissociation radius.  Very similar shapes of rotational lines of
SiO (J=2-1 through J=7-6 in their ground vibrational state) have been
observed in IRC$+$10216 with the IRAM 30-m telescope by \cite{Agundez2012} (see
their Fig.\,4), who suggested that the observed peaks may be formed
by additional excitation provided by shells with enhanced density observed
in the outer layers of the envelope of IRC$+$10216 \citep[see
e.g.][]{Cernicharo:2015uq}.  The line shape can also indicate the presence of
spiral shells \citep{Decin:2015qy} in the inner envelope, and/or a hole in
the middle of the envelope in the ammonia distribution.  

Another possible hypothesis that explains the slightly double-peaked shape of
NH$_3$ rotational transitions could be additional excitation of levels by
line overlap.  In particular, this effect may couple ortho- and para-NH$_3$
levels.  This hypothesis has been tested by simultaneous computations of
both ortho and para species including effects of line overlap.  We found,
however, that this process cannot explain the asymmetric double-peak line
profiles that are observed.  On the other hand, it does reproduce the
resulting composite profile of the overlapping transitions (para-NH$_3$
$3_1$(s)$-2_1$(a) at 1763.601 GHz and ortho-NH$_3$ 3$_0$(s)-2$_0$(a) at
1763.524 GHz) shown in Figure~\ref{figb} by the long dash line.

The theoretical profiles for the inversion transitions show a complex
structure, due to the overlap of emission in the hyperfine components.  In
particular, the profile of the para-NH$_3$ (1,1) line (see the lowest panel
in the Fig.\,\ref{figinv}) shows at least the central and the two long
wavelength components of the hyperfine multiplet.  The two narrow peaks at
$-32$ and $-20$\,km\,s$^{-1}$ caused by the overlap of the blue edge of one
component and red edge of the next one further confirms the presence of at
least one of the short wavelength components.  Observations with higher
S/N would probably reveal the shortest wavelength
component.

\section{Summary}

We present {\it Herschel}/HIFI observations obtained with high spectral
resolution for all nine rotational transitions up to the $J = 3$ levels for
ortho- and para-NH$_3$ in the envelope of the C-rich AGB star IRC$+$10216. 
Using our numerical code MOLEXCSE, which solves the non-LTE radiative
transfer in molecular lines and dust continuum, we searched for the best-fit
parameters that explain all three ortho- and six para-NH$_3$ rotational lines. 
Computations were done separately for ortho- and para-NH$_3$.  Three free
parameters were constrained by the modelling effort: the ammonia formation
radius, its photodissociation radius and the abundance.  The best fits were
obtained when infrared pumping of NH$_3$ was included, using the gas
temperature structure from \cite{Crosas1997}, which is representative of
self-consistently computed $T_{\rm gas}$ structures.  Test computations with
the gas temperature profile of \cite{deBeck2012} showed that, while the fit
obtained is slightly worse, the best-fit parameters are only weakly
sensitive to the assumed temperature structure.

We found that the best fit to the rotational lines is obtained if the
abundance of ortho- and para-NH$_{3}$ are equal to
($2.8\pm0.5$)$\times10^{-8}$ and ($3.2^{+0.7}_{-0.6})\times10^{-8}$,
respectively.  
The fit, including both rotational and inversion lines, 
gives the ortho- and para-NH$_3$ abundances
that are quite similar (within the error bars) to those obtained from the analysis
limited to the rotational lines only
(f(ortho-NH$_3$) = $(2.4\pm0.4)\times$10$^{-8}$ and
f(para-NH$_3$) = $(3.2\pm0.5)\times$10$^{-8}$). 
These values are compatible with an ortho/para ratio of one,
characteristic for the formation of ammonia at high temperatures.  The
derived abundance of ortho-NH$_3$ has
solved the long-lasting problem of the
discrepancy between the abundance derived from the lowest submillimeter
rotational line and those from radio inversion or infrared absorption
transitions.  It was found that the main process that brought the
abundances of ammonia derived from different spectral transitions close to
each other was the inclusion of the NIR radiative pumping of ammonia via the
$10\,\mu$m $\nu_2=1$ band. The average abundance of NH$_3$ derived from rotational
lines agree to within of a factor a few with those from radio inversion or
infrared absorption transitions, and we argued that this difference may not
be significant, since both methods rely on an uncertain determination of the ammonia
column density.

In addition, since the MOLEXCSE code also offers predictions for other
NH$_3$ transitions in the radio and mid-IR range, we tested whether our
best-fit models were able to explain the observations of \cite{Gong2015} and
\cite{Keady1993}.  In the case of the inversion transitions, we found that
reducing R$_{\rm ph}$ to 1.5$\times$10$^{16}$ cm gives a very good fit to
the lower para-NH$_3$ (1,1) and (2,2) inversion transitions, but with so
small a R$_{\rm ph}$ that the fit to the rotational lines of ortho-NH$_3$ is
significantly worse.  We admit, however, that our best fit model for the
rotational transitions may overpredict the envelope size as a result of
inaccuracies in modelling of the inner parts of the envelope.  In the case of the
MIR absorption lines, we found that our best-fit model reproduces the
observed profiles from \cite{Keady1993} fairly well.  Some disagreement with
the $aQ(6,6)$ line profile is most probably due to our neglect of the inner
velocity structure in the envelope.

The ammonia formation radius is not well-constrained in our approach, and the
best fitting abundances of ortho- and para-NH$_3$ were obtained at the
minimum value of the formation radius considered in our analysis, 2.5\,R$_\star$.
The best fits were obtained assuming that the outflow velocity is
constant, so we performed test computations with the gas velocity,
increasing inside the acceleration zone. We found that for $R_{\rm
f}=2.5$\,R$_\star$ this type of velocity field produces a narrow emission feature in
the centre of the highest inversion transitions that is not seen in the
observed line profiles.  To remove this feature, created in the acceleration
region, it was necessary to shift $R_{\rm f}$ above $5-10$ R$_{\star}$,
still inside the 1$\sigma$ limits obtained from fitting the line strengths.  
Nevertheless, the more realistic gas velocity
field suggests that ammonia is not formed at the stellar photosphere as had been
suggested by the formal fits to the rotational transitions,
assuming a constant gas velocity field.

A lower limit on the ammonia photodissociation radius seems to be implied by
our fit to the rotational transitions, while the maximum value for $R_{\rm
ph}$ is unconstrained.  Our value of $(2-3)\times 10^{16}$ cm seems to agree
quite well with other determinations.  However, the best fit to the both
rotational and inversional transitions does not give fully consistent
results for the ortho- (R$_{\rm ph}=(5_{-3.}^{>+1.})\times10^{16}$\,cm) and
para-NH$_3$ (R$_{\rm ph}=(1.5\pm0.2)\times10^{16}$\,cm).  Furthermore, a
distribution of NH$_3$ with the fast decrease predicted by the model of
{\cite{Decin:2010kl} does not yield a good fit to the lowest rotational
transitions of ortho-ammonia.

The two rotational lines that have the highest S/N  (1$_0$(s)-0$_0$(a)
of ortho- at 572.498 GHz and $3_1$(s)$-2_1$(a) of para-NH$_3$ at 1763.601
GHz) show two asymmetric peaks in their profile.  These can be due to an
excess of radiation from almost concentric rings or spiral shells as
suggested in literature.  We also investigated the possibility that they
result from the coupling of the ortho- and para-NH$_3$ levels by the effects
of line overlap.  The computations showed that this process cannot explain
the observed peaks, but it can successfully account for the
$3_1$(s)$-2_1$(a) at 1763.601 GHz and ortho-NH$_3$ 3$_0$(s)-2$_0$(a) line at
1763.524 GHz.

\begin{acknowledgements}

MSc and RSz acknowledge support by the National
Science Center under grant (N 203 581040).
JHe thanks the support of the NSFC research grant No. 11173056.
Funded by Chinese Aacademy Of Sciences President's International Fellowship
Initiative.
We acknowledge support from EU FP7-PEOPLE-2010-IRSES programme in the
framework of project POSTAGBinGALAXIES (Grant Agreement No.269193).  
Grant No.  2015VMA015.  
V.B., J.A. and P.P. acknowledge support by the Spanish MICINN, program CONSOLIDER
INGENIO 2010, grant 'ASTROMOL' (CSD2009-00038), and MINECO, grants
AYA2012-32032 and FIS2012-32096.
J. Cernicharo thanks Spanish MINECO for funding
under grants AYA2009-07304, AYA2012-32032, CSD2009-00038, and
ERC under ERC-2013-SyG, G.A. 610256 NANOCOSMOS.
L.~D. acknowledges funding by the European Research Council under the European
Community’s H2020 program/ERC grant agreement No.  646758 (AEROSOL).
K.~J. acknowledges the funding from the Swedish national space board.
D.~A.~N. was supported by a grant issued by NASA/JPL.

HIFI has been designed and built by a
consortium of institutes and university departments from across Europe,
Canada and the United States under the leadership of SRON Netherlands
Institute for Space Research, Groningen, The Netherlands and with major
contributions from Germany, France and the US.  Consortium members are:
Canada: CSA, U.Waterloo; France: CESR, LAB, LERMA, IRAM; Germany: KOSMA,
MPIfR, MPS; Ireland, NUI Maynooth; Italy: ASI, IFSI-INAF, Osservatorio
Astrofisico di Arcetri- INAF; Netherlands: SRON, TUD; Poland: CAMK, CBK;
Spain: Observatorio Astron\'omico Nacional (IGN), Centro de
Astrobiolog\'{\i}a (CSIC-INTA); Sweden: Chalmers University of Technology -
MC2, RSS \& GARD; Onsala Space Observatory; Swedish National Space Board,
Stockholm University - Stockholm Observatory; Switzerland: ETH Zurich, FHNW;
USA: Caltech, JPL, NHSC.  HCSS / HSpot / HIPE is a joint development by the
Herschel Science Ground Segment Consortium, consisting of ESA, the NASA
Herschel Science Center, and the HIFI, PACS and SPIRE consortia.

\end{acknowledgements}

\bibliographystyle{aa} 
\bibliography{schmidt} 

\begin{thebibliography}{63}
\expandafter\ifx\csname natexlab\endcsname\relax\def\natexlab#1{#1}\fi

\bibitem[{{Ag{\'u}ndez} \& {Cernicharo}(2006)}]{Agundez2006}
{Ag{\'u}ndez}, M. \& {Cernicharo}, J. 2006, \apj, 650, 374

\bibitem[{{Ag{\'u}ndez} {et~al.}(2010){Ag{\'u}ndez}, {Cernicharo}, \&
  {Gu{\'e}lin}}]{Agundez2010}
{Ag{\'u}ndez}, M., {Cernicharo}, J., \& {Gu{\'e}lin}, M. 2010, \apjl, 724, L133

\bibitem[{{Ag{\'u}ndez} {et~al.}(2012){Ag{\'u}ndez}, {Fonfr{\'{\i}}a},
  {Cernicharo}, {Kahane}, {Daniel}, \& {Gu{\'e}lin}}]{Agundez2012}
{Ag{\'u}ndez}, M., {Fonfr{\'{\i}}a}, J.~P., {Cernicharo}, J., {et~al.} 2012,
  \aap, 543, A48

\bibitem[{{Becklin} {et~al.}(1969){Becklin}, {Frogel}, {Hyland}, {Kristian}, \&
  {Neugebauer}}]{Becklin1969}
{Becklin}, E.~E., {Frogel}, J.~A., {Hyland}, A.~R., {Kristian}, J., \&
  {Neugebauer}, G. 1969, \apjl, 158, L133

\bibitem[{{Bell} {et~al.}(1980){Bell}, {Kwok}, \& {Feldman}}]{Bell:1980hc}
{Bell}, M.~B., {Kwok}, S., \& {Feldman}, P.~A. 1980, in IAU Symposium, Vol.~87,
  Interstellar Molecules, ed. B.~H. {Andrew}, 495

\bibitem[{{Bell} {et~al.}(1982){Bell}, {Kwok}, {Matthews}, \&
  {Feldman}}]{Bell1982}
{Bell}, M.~B., {Kwok}, S., {Matthews}, H.~E., \& {Feldman}, P.~A. 1982, \aj,
  87, 404

\bibitem[{{Betz}(1987)}]{Betz:1987lr}
{Betz}, A. 1987, in IAU Symposium, Vol. 120, Astrochemistry, ed. M.~S. {Vardya}
  \& S.~P. {Tarafdar}, 327--336

\bibitem[{{Betz} {et~al.}(1979){Betz}, {McLaren}, \& {Spears}}]{Betz1979}
{Betz}, A.~L., {McLaren}, R.~A., \& {Spears}, D.~L. 1979, \apjl, 229, L97

\bibitem[{Bunker \& Jensen(1998)}]{BunkerJensen}
Bunker, P.~R. \& Jensen, P. 1998, Molecular symmetry and spectroscopy, 2nd edn.
  (Ottawa, Canada: NRC Research Press)

\bibitem[{{Cernicharo} {et~al.}(2015){Cernicharo}, {Marcelino}, {Ag{\'u}ndez},
  \& {Gu{\'e}lin}}]{Cernicharo:2015uq}
{Cernicharo}, J., {Marcelino}, N., {Ag{\'u}ndez}, M., \& {Gu{\'e}lin}, M. 2015,
  \aap, 575, A91

\bibitem[{{Cherchneff}(2012)}]{Cherchneff2012}
{Cherchneff}, I. 2012, \aap, 545, A12

\bibitem[{{Cheung} {et~al.}(1968){Cheung}, {Rank}, {Townes}, {Thornton}, \&
  {Welch}}]{Cheung1968}
{Cheung}, A.~C., {Rank}, D.~M., {Townes}, C.~H., {Thornton}, D.~D., \& {Welch},
  W.~J. 1968, Physical Review Letters, 21, 1701

\bibitem[{{Crosas} \& {Menten}(1997)}]{Crosas1997}
{Crosas}, M. \& {Menten}, K.~M. 1997, \apj, 483, 913

\bibitem[{{Danby} {et~al.}(1988){Danby}, {Flower}, {Valiron}, {Schilke}, \&
  {Walmsley}}]{Danby1988}
{Danby}, G., {Flower}, D.~R., {Valiron}, P., {Schilke}, P., \& {Walmsley},
  C.~M. 1988, \mnras, 235, 229

\bibitem[{{Danilovich} {et~al.}(2014){Danilovich}, {Bergman}, {Justtanont},
  {Lombaert}, {Maercker}, {Olofsson}, {Ramstedt}, \& {Royer}}]{Danilovich2014}
{Danilovich}, T., {Bergman}, P., {Justtanont}, K., {et~al.} 2014, \aap, 569,
  A76

\bibitem[{{De Beck} {et~al.}(2012){De Beck}, {Lombaert}, {Ag{\'u}ndez},
  {Daniel}, {Decin}, {Cernicharo}, {M{\"u}ller}, {Min}, {Royer},
  {Vandenbussche}, {de Koter}, {Waters}, {Groenewegen}, {Barlow}, {Gu{\'e}lin},
  {Kahane}, {Pearson}, {Encrenaz}, {Szczerba}, \& {Schmidt}}]{deBeck2012}
{De Beck}, E., {Lombaert}, R., {Ag{\'u}ndez}, M., {et~al.} 2012, \aap, 539,
  A108

\bibitem[{{de Graauw} {et~al.}(2010){de Graauw}, {Helmich}, {Phillips},
  {Stutzki}, {Caux}, {Whyborn}, {Dieleman}, {Roelfsema}, {Aarts}, {Assendorp},
  {Bachiller}, {Baechtold}, {Barcia}, {Beintema}, {Belitsky}, {Benz}, {Bieber},
  {Boogert}, {Borys}, {Bumble}, {Ca{\"\i}s}, {Caris}, {Cerulli-Irelli},
  {Chattopadhyay}, {Cherednichenko}, {Ciechanowicz}, {Coeur-Joly}, {Comito},
  {Cros}, {de Jonge}, {de Lange}, {Delforges}, {Delorme}, {den Boggende},
  {Desbat}, {Diez-Gonz{\'a}lez}, {di Giorgio}, {Dubbeldam}, {Edwards},
  {Eggens}, {Erickson}, {Evers}, {Fich}, {Finn}, {Franke}, {Gaier}, {Gal},
  {Gao}, {Gallego}, {Gauffre}, {Gill}, {Glenz}, {Golstein}, {Goulooze},
  {Gunsing}, {G{\"u}sten}, {Hartogh}, {Hatch}, {Higgins}, {Honingh}, {Huisman},
  {Jackson}, {Jacobs}, {Jacobs}, {Jarchow}, {Javadi}, {Jellema}, {Justen},
  {Karpov}, {Kasemann}, {Kawamura}, {Keizer}, {Kester}, {Klapwijk}, {Klein},
  {Kollberg}, {Kooi}, {Kooiman}, {Kopf}, {Krause}, {Krieg}, {Kramer},
  {Kruizenga}, {Kuhn}, {Laauwen}, {Lai}, {Larsson}, {Leduc}, {Leinz}, {Lin},
  {Liseau}, {Liu}, {Loose}, {L{\'o}pez-Fernandez}, {Lord}, {Luinge}, {Marston},
  {Mart{\'{\i}}n-Pintado}, {Maestrini}, {Maiwald}, {McCoey}, {Mehdi}, {Megej},
  {Melchior}, {Meinsma}, {Merkel}, {Michalska}, {Monstein}, {Moratschke},
  {Morris}, {Muller}, {Murphy}, {Naber}, {Natale}, {Nowosielski}, {Nuzzolo},
  {Olberg}, {Olbrich}, {Orfei}, {Orleanski}, {Ossenkopf}, {Peacock}, {Pearson},
  {Peron}, {Phillip-May}, {Piazzo}, {Planesas}, {Rataj}, {Ravera}, {Risacher},
  {Salez}, {Samoska}, {Saraceno}, {Schieder}, {Schlecht}, {Schl{\"o}der},
  {Schm{\"u}lling}, {Schultz}, {Schuster}, {Siebertz}, {Smit}, {Szczerba},
  {Shipman}, {Steinmetz}, {Stern}, {Stokroos}, {Teipen}, {Teyssier}, {Tils},
  {Trappe}, {van Baaren}, {van Leeuwen}, {van de Stadt}, {Visser}, {Wildeman},
  {Wafelbakker}, {Ward}, {Wesselius}, {Wild}, {Wulff}, {Wunsch}, {Tielens},
  {Zaal}, {Zirath}, {Zmuidzinas}, \& {Zwart}}]{de-Graauw:2010qe}
{de Graauw}, T., {Helmich}, F.~P., {Phillips}, T.~G., {et~al.} 2010, \aap, 518,
  L6

\bibitem[{{Decin} {et~al.}(2010){Decin}, {Ag{\'u}ndez}, {Barlow}, {Daniel},
  {Cernicharo}, {Lombaert}, {De Beck}, {Royer}, {Vandenbussche}, {Wesson},
  {Polehampton}, {Blommaert}, {De Meester}, {Exter}, {Feuchtgruber}, {Gear},
  {Gomez}, {Groenewegen}, {Gu{\'e}lin}, {Hargrave}, {Huygen}, {Imhof},
  {Ivison}, {Jean}, {Kahane}, {Kerschbaum}, {Leeks}, {Lim}, {Matsuura},
  {Olofsson}, {Posch}, {Regibo}, {Savini}, {Sibthorpe}, {Swinyard}, {Yates}, \&
  {Waelkens}}]{Decin:2010kl}
{Decin}, L., {Ag{\'u}ndez}, M., {Barlow}, M.~J., {et~al.} 2010, \nat, 467, 64

\bibitem[{{Decin} {et~al.}(2015){Decin}, {Richards}, {Neufeld}, {Steffen},
  {Melnick}, \& {Lombaert}}]{Decin:2015qy}
{Decin}, L., {Richards}, A.~M.~S., {Neufeld}, D., {et~al.} 2015, \aap, 574, A5

\bibitem[{{Gong} {et~al.}(2015){Gong}, {Henkel}, {Spezzano}, {Thorwirth},
  {Menten}, {Wyrowski}, {Mao}, \& {Klein}}]{Gong2015}
{Gong}, Y., {Henkel}, C., {Spezzano}, S., {et~al.} 2015, \aap, 574, A56

\bibitem[{{Groenewegen} {et~al.}(2012){Groenewegen}, {Barlow}, {Blommaert},
  {Cernicharo}, {Decin}, {Gomez}, {Hargrave}, {Kerschbaum}, {Ladjal}, {Lim},
  {Matsuura}, {Olofsson}, {Sibthorpe}, {Swinyard}, {Ueta}, \&
  {Yates}}]{Groenewegen:2012fu}
{Groenewegen}, M.~A.~T., {Barlow}, M.~J., {Blommaert}, J.~A.~D.~L., {et~al.}
  2012, \aap, 543, L8

\bibitem[{{Groenewegen} {et~al.}(1998){Groenewegen}, {van der Veen}, \&
  {Matthews}}]{Groenewegen:1998rw}
{Groenewegen}, M.~A.~T., {van der Veen}, W.~E.~C.~J., \& {Matthews}, H.~E.
  1998, \aap, 338, 491

\bibitem[{{Harju} {et~al.}(1993){Harju}, {Walmsley}, \&
  {Wouterloot}}]{Harju1993}
{Harju}, J., {Walmsley}, C.~M., \& {Wouterloot}, J.~G.~A. 1993, \aaps, 98, 51

\bibitem[{{Hasegawa} {et~al.}(2006){Hasegawa}, {Kwok}, {Koning}, {Volk},
  {Justtanont}, {Olofsson}, {Sch{\"o}ier}, {Sandqvist}, {Hjalmarson}, {Olberg},
  {Winnberg}, {Nyman}, \& {Frisk}}]{Hasegawa2006}
{Hasegawa}, T.~I., {Kwok}, S., {Koning}, N., {et~al.} 2006, \apj, 637, 791

\bibitem[{{Herbig} \& {Zappala}(1970)}]{Herbig1970}
{Herbig}, G.~H. \& {Zappala}, R.~R. 1970, \apjl, 162, L15

\bibitem[{{Ho} \& {Townes}(1983)}]{Ho1983}
{Ho}, P.~T.~P. \& {Townes}, C.~H. 1983, \araa, 21, 239

\bibitem[{{Jijina} {et~al.}(1999){Jijina}, {Myers}, \& {Adams}}]{Jijina1999}
{Jijina}, J., {Myers}, P.~C., \& {Adams}, F.~C. 1999, \apjs, 125, 161

\bibitem[{{Keady} \& {Ridgway}(1993)}]{Keady1993}
{Keady}, J.~J. \& {Ridgway}, S.~T. 1993, \apj, 406, 199

\bibitem[{{Kirsanova} {et~al.}(2014){Kirsanova}, {Wiebe}, {Sobolev}, {Henkel},
  \& {Tsivilev}}]{Kirsanova2014}
{Kirsanova}, M.~S., {Wiebe}, D.~S., {Sobolev}, A.~M., {Henkel}, C., \&
  {Tsivilev}, A.~P. 2014, \mnras, 437, 1593

\bibitem[{{Kukolich}(1967)}]{Kukolich1967}
{Kukolich}, S.~G. 1967, Phys.~Rev., 156, 83

\bibitem[{{Kwok} {et~al.}(1981){Kwok}, {Bell}, \& {Feldman}}]{Kwok1981}
{Kwok}, S., {Bell}, M.~B., \& {Feldman}, P.~A. 1981, \apj, 247, 125

\bibitem[{{Le Bertre}(1992)}]{LeBertre1992}
{Le Bertre}, T. 1992, \aaps, 94, 377

\bibitem[{{Mamon} {et~al.}(1988){Mamon}, {Glassgold}, \& {Huggins}}]{Mamon1988}
{Mamon}, G.~A., {Glassgold}, A.~E., \& {Huggins}, P.~J. 1988, \apj, 328, 797

\bibitem[{{McLaren} \& {Betz}(1980)}]{McLaren1980}
{McLaren}, R.~A. \& {Betz}, A.~L. 1980, \apjl, 240, L159

\bibitem[{{Men'shchikov} {et~al.}(2001){Men'shchikov}, {Balega}, {Bl{\"o}cker},
  {Osterbart}, \& {Weigelt}}]{Menshchikov:2001kl}
{Men'shchikov}, A.~B., {Balega}, Y., {Bl{\"o}cker}, T., {Osterbart}, R., \&
  {Weigelt}, G. 2001, \aap, 368, 497

\bibitem[{{Menten} {et~al.}(2012){Menten}, {Reid}, {Kami{\'n}ski}, \&
  {Claussen}}]{Menten2012}
{Menten}, K.~M., {Reid}, M.~J., {Kami{\'n}ski}, T., \& {Claussen}, M.~J. 2012,
  \aap, 543, A73

\bibitem[{{Menten} {et~al.}(2010){Menten}, {Wyrowski}, {Alcolea}, {De Beck},
  {Decin}, {Marston}, {Bujarrabal}, {Cernicharo}, {Dominik}, {Justtanont}, {de
  Koter}, {Melnick}, {Neufeld}, {Olofsson}, {Planesas}, {Schmidt},
  {Sch{\"o}ier}, {Szczerba}, {Teyssier}, {Waters}, {Edwards}, {Olberg},
  {Phillips}, {Morris}, {Salez}, \& {Caux}}]{Menten:2010pd}
{Menten}, K.~M., {Wyrowski}, F., {Alcolea}, J., {et~al.} 2010, \aap, 521, L7

\bibitem[{{Mihalas} {et~al.}(1975){Mihalas}, {Kunasz}, \&
  {Hummer}}]{Mihalas1975}
{Mihalas}, D., {Kunasz}, P.~B., \& {Hummer}, D.~G. 1975, \apj, 202, 465

\bibitem[{{Miller}(1970)}]{Miller1970}
{Miller}, J.~S. 1970, \apjl, 161, L95

\bibitem[{{Monnier} {et~al.}(2000){Monnier}, {Danchi}, {Hale}, {Tuthill}, \&
  {Townes}}]{Monnier2000b}
{Monnier}, J.~D., {Danchi}, W.~C., {Hale}, D.~S., {Tuthill}, P.~G., \&
  {Townes}, C.~H. 2000, \apj, 543, 868

\bibitem[{{Neufeld} {et~al.}(2010){Neufeld}, {Gonz{\'a}lez-Alfonso}, {Melnick},
  {Pu{\l}ecka}, {Schmidt}, {Szczerba}, {Bujarrabal}, {Alcolea}, {Cernicharo},
  {Decin}, {Dominik}, {Justtanont}, {de Koter}, {Marston}, {Menten},
  {Olofsson}, {Planesas}, {Sch{\"o}ier}, {Teyssier}, {Waters}, {Edwards},
  {McCoey}, {Shipman}, {Jellema}, {de Graauw}, {Ossenkopf}, {Schieder}, \&
  {Philipp}}]{Neufeld2010}
{Neufeld}, D.~A., {Gonz{\'a}lez-Alfonso}, E., {Melnick}, G., {et~al.} 2010,
  \aap, 521, L5

\bibitem[{{Neufeld} {et~al.}(2013){Neufeld}, {Tolls}, {Ag{\'u}ndez},
  {Gonz{\'a}lez-Alfonso}, {Decin}, {Daniel}, {Cernicharo}, {Melnick},
  {Schmidt}, \& {Szczerba}}]{Neufeld:2013fk}
{Neufeld}, D.~A., {Tolls}, V., {Ag{\'u}ndez}, M., {et~al.} 2013, \apjl, 767, L3

\bibitem[{{Nguyen-Q-Rieu} {et~al.}(1984){Nguyen-Q-Rieu}, {Graham}, \&
  {Bujarrabal}}]{Nguyen-Q-Rieu1984}
{Nguyen-Q-Rieu}, {Graham}, D., \& {Bujarrabal}, V. 1984, \aap, 138, L5

\bibitem[{{Nguyen-Q-Rieu} {et~al.}(1986){Nguyen-Q-Rieu}, {Winnberg}, \&
  {Bujarrabal}}]{Nguyen-Q-Rieu:1986th}
{Nguyen-Q-Rieu}, {Winnberg}, A., \& {Bujarrabal}, V. 1986, \aap, 165, 204

\bibitem[{{Osorio} {et~al.}(2009){Osorio}, {Anglada}, {Lizano}, \&
  {D'Alessio}}]{Osorio2009}
{Osorio}, M., {Anglada}, G., {Lizano}, S., \& {D'Alessio}, P. 2009, \apj, 694,
  29

\bibitem[{{Pegourie}(1988)}]{Pegourie:1988eu}
{Pegourie}, B. 1988, \aap, 194, 335

\bibitem[{{Pilbratt} {et~al.}(2010){Pilbratt}, {Riedinger}, {Passvogel},
  {Crone}, {Doyle}, {Gageur}, {Heras}, {Jewell}, {Metcalfe}, {Ott}, \&
  {Schmidt}}]{Pilbratt:2010oq}
{Pilbratt}, G.~L., {Riedinger}, J.~R., {Passvogel}, T., {et~al.} 2010, \aap,
  518, L1

\bibitem[{{Roelfsema} {et~al.}(2012){Roelfsema}, {Helmich}, {Teyssier},
  {Ossenkopf}, {Morris}, {Olberg}, {Shipman}, {Risacher}, {Akyilmaz},
  {Assendorp}, {Avruch}, {Beintema}, {Biver}, {Boogert}, {Borys}, {Braine},
  {Caris}, {Caux}, {Cernicharo}, {Coeur-Joly}, {Comito}, {de Lange},
  {Delforge}, {Dieleman}, {Dubbeldam}, {de Graauw}, {Edwards}, {Fich},
  {Flederus}, {Gal}, {di Giorgio}, {Herpin}, {Higgins}, {Hoac}, {Huisman},
  {Jarchow}, {Jellema}, {de Jonge}, {Kester}, {Klein}, {Kooi}, {Kramer},
  {Laauwen}, {Larsson}, {Leinz}, {Lord}, {Lorenzani}, {Luinge}, {Marston},
  {Mart{\'{\i}}n-Pintado}, {McCoey}, {Melchior}, {Michalska}, {Moreno},
  {M{\"u}ller}, {Nowosielski}, {Okada}, {Orlea{\'n}ski}, {Phillips}, {Pearson},
  {Rabois}, {Ravera}, {Rector}, {Rengel}, {Sagawa}, {Salomons},
  {S{\'a}nchez-Su{\'a}rez}, {Schieder}, {Schl{\"o}der}, {Schm{\"u}lling},
  {Soldati}, {Stutzki}, {Thomas}, {Tielens}, {Vastel}, {Wildeman}, {Xie},
  {Xilouris}, {Wafelbakker}, {Whyborn}, {Zaal}, {Bell}, {Bjerkeli}, {De Beck},
  {Cavali{\'e}}, {Crockett}, {Hily-Blant}, {Kama}, {Kaminski}, {Lefl{\'o}ch},
  {Lombaert}, {de Luca}, {Makai}, {Marseille}, {Nagy}, {Pacheco}, {van der
  Wiel}, {Wang}, \& {Y{\i}ld{\i}z}}]{Roelfsema2012}
{Roelfsema}, P.~R., {Helmich}, F.~P., {Teyssier}, D., {et~al.} 2012, \aap, 537,
  A17

\bibitem[{{Rouleau} \& {Martin}(1991)}]{Rouleau:1991nx}
{Rouleau}, F. \& {Martin}, P.~G. 1991, \apj, 377, 526

\bibitem[{{Rydbeck} {et~al.}(1977){Rydbeck}, {Sume}, {Hjalmarson}, {Eld{\'e}r},
  {R{\"o}nn{\"a}ng}, \& {Kolberg}}]{Rydbeck1977}
{Rydbeck}, O.~E.~H., {Sume}, A., {Hjalmarson}, A., {et~al.} 1977, \apjl, 215,
  35

\bibitem[{{Schoenberg} \& {Hempe}(1986)}]{Schoenberg1986}
{Schoenberg}, K. \& {Hempe}, K. 1986, \aap, 163, 151

\bibitem[{{Sch{\"o}ier} {et~al.}(2011){Sch{\"o}ier}, {Maercker}, {Justtanont},
  {Olofsson}, {Black}, {Decin}, {de Koter}, \& {Waters}}]{Schoier2011}
{Sch{\"o}ier}, F.~L., {Maercker}, M., {Justtanont}, K., {et~al.} 2011, \aap,
  530, A83

\bibitem[{{Sch{\"o}ier} \& {Olofsson}(2001)}]{Schoier2001}
{Sch{\"o}ier}, F.~L. \& {Olofsson}, H. 2001, \aap, 368, 969

\bibitem[{{Sch{\"o}ier} {et~al.}(2002){Sch{\"o}ier}, {Ryde}, \&
  {Olofsson}}]{Schoier:2002nx}
{Sch{\"o}ier}, F.~L., {Ryde}, N., \& {Olofsson}, H. 2002, \aap, 391, 577

\bibitem[{{Sch{\"o}ier} {et~al.}(2005){Sch{\"o}ier}, {van der Tak}, {van
  Dishoeck}, \& {Black}}]{Schoier2005}
{Sch{\"o}ier}, F.~L., {van der Tak}, F.~F.~S., {van Dishoeck}, E.~F., \&
  {Black}, J.~H. 2005, \aap, 432, 369

\bibitem[{{Szczerba} {et~al.}(1997){Szczerba}, {Omont}, {Volk}, {Cox}, \&
  {Kwok}}]{Szczerba:1997eu}
{Szczerba}, R., {Omont}, A., {Volk}, K., {Cox}, P., \& {Kwok}, S. 1997, \aap,
  317, 859

\bibitem[{{Thaddeus} {et~al.}(1964){Thaddeus}, {Krisher}, \&
  {Loubser}}]{Thaddeus1964}
{Thaddeus}, P., {Krisher}, L.~C., \& {Loubser}, J.~H.~N. 1964, \jcp, 40, 257

\bibitem[{{Truong-Bach} {et~al.}(1991){Truong-Bach}, {Morris}, \&
  {Nguyen-Q-Rieu}}]{Truong-Bach:1991qa}
{Truong-Bach}, {Morris}, D., \& {Nguyen-Q-Rieu}. 1991, \aap, 249, 435

\bibitem[{{Truong-Bach} {et~al.}(1987){Truong-Bach}, {Nguyen-Q-Rieu}, {Omont},
  {Olofsson}, \& {Johansson}}]{TruongBach1987}
{Truong-Bach}, {Nguyen-Q-Rieu}, {Omont}, A., {Olofsson}, H., \& {Johansson},
  L.~E.~B. 1987, \aap, 176, 285

\bibitem[{{Umemoto} {et~al.}(1999){Umemoto}, {Mikami}, {Yamamoto}, \&
  {Hirano}}]{Umemoto:1999bs}
{Umemoto}, T., {Mikami}, H., {Yamamoto}, S., \& {Hirano}, N. 1999, \apjl, 525,
  L105

\bibitem[{{Walmsley} \& {Ungerechts}(1983)}]{Walmsley1983}
{Walmsley}, C.~M. \& {Ungerechts}, H. 1983, \aap, 122, 164

\bibitem[{{Wienen} {et~al.}(2012){Wienen}, {Wyrowski}, {Schuller}, {Menten},
  {Walmsley}, {Bronfman}, \& {Motte}}]{Wienen2012}
{Wienen}, M., {Wyrowski}, F., {Schuller}, F., {et~al.} 2012, \aap, 544, A146

\bibitem[{{Yurchenko} {et~al.}(2011){Yurchenko}, {Barber}, \&
  {Tennyson}}]{Yurchenko2011}
{Yurchenko}, S.~N., {Barber}, R.~J., \& {Tennyson}, J. 2011, \mnras, 413, 1828

\end{thebibliography}

\end{document}